\documentclass{article}%
\usepackage{makeidx}
\usepackage{graphicx}
\usepackage{amsmath}
\usepackage{amsfonts}
\usepackage{amssymb}%
\providecommand{\U}[1]{\protect\rule{.1in}{.1in}}

\makeindex
\begin{document}

\title{Evolutionary stability in quantum games}
\author{Azhar Iqbal\thanks{Permanent address: Centre for Advanced Mathematics and
Physics, National University of Sciences \& Technology, Campus of College of
Electrical and Mechanical Engineering, Peshawar Road, Rawalpindi, Pakistan.}
and Taksu Cheon\\Kochi University of Technology\\Tosa Yamada, Kochi 782-8502, Japan.}
\maketitle
\tableofcontents

\begin{abstract}
In evolutionary game theory an Evolutionarily Stable Strategy (ESS) is a
refinement of the Nash equilibrium concept that is sometimes also recognized
as evolutionary stability. It is a game-theoretic model, well known to
mathematical biologists, that was found quite useful in the understanding of
evolutionary dynamics of a population. This chapter presents an analysis of
evolutionary stability in the emerging field of quantum games.

\end{abstract}

\section{Introduction}

Games such as chess, warfare and politics have been played throughout history.
Whenever individuals meet who have conflicting desires, and priorities, then
games are likely to be played. Analysis and understanding of games has existed
for long times but the emergence of game theory as a formal study of games is
widely believed to have taken place when Neumann
\index{Neumann v. J}
and Morgenstern
\index{Morgenstern O.}
\cite{Neumann} published their pioneering book ``The Theory of Games and
Economic Behaviour'' in 1944. Game theory \cite{Rasmusen89}\ is now an
established discipline of mathematics that is a vast subject having a rich
history and content. Roughly speaking, game theory is the analysis of the
actions made by rational players when these actions are strategically interdependent.

The 1970s saw game theory
\index{Game theory}
being successfully applied to problems of evolutionary biology and a new
branch of game theory, recognized as evolutionary game theory
\index{Evolutionary game theory}
\cite{Smith,Hofbauer,Weibull}, came into existence. The concept of utility
from economics was given an interpretation in terms of Darwinian fitness
\index{Darwinian fitness}%
. Originally, evolutionary game theory
\index{Evolutionary game theory}
considered animal conflicts occurring in macro-world. In recent years,
research in biology \cite{Turner} suggested that nature also plays classical
games at micro-level. Bacterial infections by viruses are classical game-like
situations where nature prefers dominant strategies
\index{Dominant strategy}%
.

In game theory \cite{Neumann,Rasmusen89} one finds many examples where
multiple Nash equilibria
\index{Nash equilibrium}
(NE) \cite{JohnNash,JohnNash1} emerge as solutions of a game. To select one
(or possibly more) out of these requires some refinement of the equilibrium
\index{Refinement of Nash equilibrium}
concept \cite{MyersonRB}. A refinement is a rule/criterion that describes the
criterion to prefer one (in some cases more than one) equilibrium out of many.
Numerous refinements are found in game theory, for example, perfect
equilibrium
\index{Perfect equilibrium}
(used for extensive- and normal-form games), sequential equilibrium
\index{Sequential equilibrium}
(a fundamental non-cooperative solution concept for extensive-form games), and
correlated equilibrium
\index{Correlated equilibrium}
(used for modelling communication among players).

During recent years quantum game theory
\index{Quantum game theory}
\cite{MeyerDavid,Eisert,Eisert1} has emerged as a new research field within
quantum information and computation
\index{Quantum information and computation}
\cite{Nielsen}. A significant portion of research in quantum games
\index{Quantum games}
deals with the question asking how quantization of a game
\index{Quantization of a game}
affects/changes the existence/location of a NE. This question has been
addressed in a number of publications \cite{PhDThesises} in this area and now
it seems that it is generally agreed that quantization of a game
\index{Quantization of a game}
indeed affects/changes the existence/location of a NE.

In this chapter we argue that, like asking how quantization of a game
\index{Quantization of a game}
affects/changes the existence/location of a NE, an equally important question
for quantum games
\index{Quantum games}
is to ask how quantization of a game
\index{Quantization of a game}
can affect a refinement of the NE
\index{Refinement of Nash equilibrium}
concept. We notice that a particular refinement of the NE, known as an
Evolutionarily Stable Strategy
\index{Evolutionarily Stable Strategy}
(ESS
\index{ESS}%
), is central to evolutionary game theory
\index{Evolutionary game theory}%
. While focussing on a this refinement, we motivate those quantum games
\index{Quantum games}
in which a NE persists\footnote{By saying that a NE persists in both the
classical and quantum version of a game we mean that there exists a NE
consisting of quantum strategies that rewards both the players exactly the
same the corresponding NE does in the classical version of the game.} in both
of its classical and quantum versions while its property of being an ESS
survives in either classical or its quantum version, but not in the both. We
argue that, the quantum games
\index{Quantum games}
offering such situations, along with their quantization procedures, can
justifiably be said to extend the boundary of investigations in quantum games
\index{Quantum games}
from existence/location of NE to existence/location of one (or more) of its refinements.

\section{Evolutionary game theory and evolutionary stability
\index{Evolutionary stability}%
}

The roots of evolutionary game theory
\index{Evolutionary game theory}
\cite{Weibull} can be traced to the puzzle of the approximate equality of the
sex ratio
\index{Sex ratio}
in mammals. In 1930 Fisher
\index{Fisher}
( \cite{Fisher,Stanford}) noticed that if individual fitness is defined in
terms of the expected number of grandchildren, then it becomes dependent on
how males and females are distributed in a population. Fisher
\index{Fisher}
showed that the evolutionary dynamics
\index{Evolutionary dynamics}
then leads to the sex ratio becoming fixed at equal numbers of males and
females. Although Fisher's argument
\index{Fisher}
can be recast in game-theoretic language but originally it was not presented
in those terms. Perhaps it was due to the fact that until that time modern
game theory had not yet emerged as a formal study of games.

Modern game theory was used, for the first time, to understand evolution when
in 1972 Maynard Smith
\index{Maynard Smith}
and G. R. Price
\index{Price G. R.}
introduced the concept of an Evolutionarily Stable Strategy (ESS) \cite{Smith
Price,Smith}. Presently, this concept is widely believed to be the cornerstone
of evolutionary game theory
\index{Evolutionary game theory}
\cite{Hofbauer} and has been found quite useful to explain and understand
animal behavior.

Traditionally, game theory
\index{Game theory}
had concerned analyzing interactions among hyperrational players and the idea
that it can be applied to animals seemed strange at the time. The ESS concept
\index{ESS}
made three important changes in the traditional meaning of the concepts of a)
strategy, b) equilibrium, and c) players' interactions.

a) \emph{Strategy}
\index{Strategy}%
: In traditional game theory, players have strategy set
\index{Strategy set}
from which they choose their strategies. In biology, animals belonging to a
species have strategy sets that are genotypic variants that may be mutated, of
which individuals inherit one or another variant, which they then play in
their strategic interactions. A mixed strategy in game theory means a convex
linear combination (with real and normalized coefficients) of pure strategies.
Because genotypic variants are taken as pure strategies, the evolutionary game
theory interprets a mixed strategy in terms of proportion of the population
that is playing that strategy.

b) \emph{Equilibrium}
\index{Equilibrium}%
: An ESS represents an equilibrium and it is a strategy
\index{Strategy}
having the property that if a whole population plays it, it cannot be invaded
under the influence of natural selection
\index{Natural selection}%
, by a small group of players playing mutant strategies. Because strategies of
evolutionary games are genotypes
\index{Genotype}
the ESS definition takes the following form: If adapted by a whole population
an ESS is a genotype that cannot be invaded by another genotype when it
appears in a small fraction of the total population.

c) \emph{Player interactions}
\index{Player interaction}%
: The ESS concept is about repeated and random pairing of players who play
strategies based on their genome
\index{Genome}
and \emph{not} on the previous history of play. This approach was new to the
usual approach of one-shot and repeated games of classical game theory
\index{Classical game theory}%
.

Consider a large population \cite{Weibull,Hofbauer} in which members are
matched repeatedly and randomly in pairs to play a bi-matrix game
\index{Bi-matrix game}%
. The players are anonymous, that is, any pair of players plays the same
symmetric bi-matrix game
\index{Bi-matrix game}%
. The symmetry of a bi-matrix game
\index{Symmetric game}
means that for a strategy set $S$ Alice's payoff when she plays $S_{1}\in S$
and Bob plays $S_{2}\in S$ is the same as Bob's payoff when he plays $S_{1}$
and Alice plays $S_{2}$. Hence, a player's payoff is defined by his/her
strategy
\index{Strategy}
and \emph{not} by his/her identity and an exchange of strategies by the two
players also exchanges their respective payoffs. A symmetric bi-matrix game
\index{Bi-matrix game}
is represented by an expression $G=(M,M^{T})$ where $M $ is the first player's
payoff matrix
\index{Payoff matrix}
and $M^{T}$, being its transpose, is the second players' payoff matrix
\index{Payoff matrix}%
. In a symmetric pair-wise contest
\index{Pair-wise contest}
one can write $P(x,y)$ as being the payoff to a $x$-player against a $y$-player.

To be precise \cite{Hofbauer,Bomze,Bomze1} a strategy
\index{Strategy}
$x$ is said to be an ESS if:

a) for each mutant strategy
\index{Mutant strategy}
$y$ there exists a positive \textit{invasion barrier}
\index{Invasion barrier}%
\textit{.}

b) if the population share of individuals playing the mutant strategy
\index{Mutant strategy}
$y$ falls below the invasion barrier, then $x$ earns a higher expected payoff
than $y$.

Mathematically speaking \cite{Weibull,Hofbauer} $x$ is an ESS when for each
strategy
\index{Strategy}
$y\neq x$ the inequality%

\begin{equation}
P[x,(1-\epsilon)x+\epsilon y]>P[y,(1-\epsilon)x+\epsilon y] \label{ESSDefIneq}%
\end{equation}
holds for all sufficiently small $\epsilon>0$. In (\ref{ESSDefIneq}) the
expression on the left-hand side is payoff to the strategy $x$ when played
against the mixed strategy
\index{Mixed strategy}
$(1-\epsilon)x+\epsilon y$. This condition for an ESS
\index{ESS}
is shown \cite{Smith Price,Smith,Weibull} equivalent to the following requirements:%

\begin{align}
\text{a) }P(x,x)  &  >P(y,x)\nonumber\\
\text{b) If\ }P(x,x)  &  =P(y,x)\text{ then}\ P(x,y)>P(y,y)\text{.}
\label{ESSDef}%
\end{align}
It turns out \cite{Smith,Weibull} that an ESS
\index{ESS}
is a symmetric NE that is stable against small mutations. Condition a) in the
definition (\ref{ESSDef}) shows $(x,x)$ is a NE for the bi-matrix game
\index{Bi-matrix game}
$G=(M,M^{T})$ if $x$ is an ESS
\index{ESS}%
. However, the converse is not true. That is, if $(x,x)$ is a NE then $x$ is
an ESS only if $x$ satisfies condition b) in definition (\ref{ESSDef}).

Evolutionary game theory
\index{Evolutionary game theory}
defines the concept of \textit{fitness} \cite{Prestwich} of a strategy
\index{Fitness of a strategy}
as follows. Suppose $x$ and $y$ are pure strategies played by a population of
players that is engaged in a two-player game. Their fitnesses are%

\begin{equation}
W(x)=P(x,x)F_{x}+P(x,y)F_{y};\text{ \ \ }W(y)=P(y,x)F_{x}+P(y,y)F_{y}
\label{fitnesses}%
\end{equation}
where $F_{x}$ and $F_{y}$\ are frequencies (the relative proportions) of the
pure strategies $x$ and $y$ respectively.

It turned out that an ESS
\index{ESS}
is a refinement on the set of symmetric Nash equilibria
\cite{Weibull,Cressman}. For symmetric bi-matrix games
\index{Bi-matrix game}
this relationship is described \cite{Gerard van} as $\bigtriangleup
^{ESS}\subset\bigtriangleup^{PE}\subset\bigtriangleup^{NE}$ where
$\bigtriangleup^{PE}\neq\Phi$ and $\bigtriangleup^{NE}$, $\bigtriangleup^{PE}%
$, $\bigtriangleup^{ESS}$ are the sets of symmetric NE, symmetric proper
equilibrium, and ESSs respectively.

The property of an ESS
\index{ESS}
of being robust against small mutations is also referred to as
\emph{evolutionary stability}
\index{Evolutionary stability}%
\emph{\ }\cite{Bomze,Bomze1}. This concept provided a significant part of the
motivation for later developments in evolutionary game theory
\index{Evolutionary game theory}%
. In evolutionary game theory, the Darwinian natural selection
\index{Darwinian natural selection}
\index{Natural selection}
is formulated as an algorithm called \emph{replicator dynamics}
\index{Replicator dynamics}
\cite{Hofbauer,Weibull} which is a mathematical statement saying that in a
population the proportion of players playing better strategies increases with
time. Mathematically, ESSs
\index{ESS}
come out as the \emph{rest points} of replicator dynamics
\index{Rest points of replicator dynamics}
\cite{Hofbauer,Weibull}.

Evolutionary stability
\index{Evolutionary stability}
was found to be a useful concept because it says something about the dynamic
properties of a system without being committed to a particular dynamic model.
Sometimes, it is also described as a model of rationality which is physically
grounded in natural selection
\index{Natural selection}%
.

\subsection{Population setting of evolutionary game theory}

Evolutionary game theory introduces so-called the \emph{population setting}
\index{Population setting}
\cite{Weibull,Hofbauer}\emph{\ }that is also known as
\emph{population-statistical setting}
\index{Population-statistical setting}%
. This setting assumes a) an infinite population of players who are engaged in
random pair-wise contests
\index{Pair-wise contest}
b) each player being programmed to play only one strategy and c) an
evolutionary pressure ensuring that better-performing strategies have better
chances of survival at the expense of other competing strategies. Because of
b) one can refer to better-performing players as better-performing strategies.

The population setting of evolutionary game theory is not alien to the concept
of the NE, although it may give such an impression. In fact, John Nash
\index{John Nash}
himself had this setting in his mind when he introduced this concept in game
theory. In his unpublished Ph.D. thesis
\index{Nash's PhD thesis}
\cite{NashThesis,Hofbauer} he wrote `\textit{it is unnecessary to assume that
the participants have...the ability to go through any complex reasoning
process. But the participants are supposed to accumulate empirical information
on the various pure strategies at their disposal...We assume that there is a
population...of participants...and that there is a stable average frequency
with which a pure strategy is employed by the ``average member'' of the
appropriate population}'.

That is, Nash had suggested a population interpretation of the NE concept in
which players are randomly drawn from large populations. Nash assumed that
these players were not aware of the total structure of the game and did not
have either the ability nor inclination to go through any complex reasoning process.

\section{Quantum games
\index{Quantum games}%
}

This chapter considers evolutionary stability
\index{Evolutionary stability}
in quantum games
\index{Quantum games}
that are played in the two quantization schemes: Eisert, Wilkens, Lewenstein
(EWL) scheme
\index{EWL quantization scheme}
\cite{Eisert,Eisert1} for playing quantum Prisoners' Dilemma
\index{Prisoners' Dilemma (PD)}
(PD) and Marinatto and Weber (MW) scheme
\index{MW quantization scheme}
\cite{Marinatto1} for playing quantum Battle of Sexes (BoS)
\index{Battle of Sexes (BoS)}
game.

EWL quantization scheme
\index{EWL quantization scheme}
appeared soon after Meyer's publication \cite{MeyerDavid} of the PQ
penny-flip
\index{PQ penny flip game}
-- a quantum game that generated significant interest and is widely believed
to have led to the creation of the new research field of quantum games
\index{Quantum games}%
. MW scheme
\index{MW quantization scheme}
derives from EWL scheme
\index{EWL quantization scheme}
but it gives a different meaning to the term `strategy'
\cite{Benjamin2,Marinatto's reply}.

EWL
\index{EWL quantization scheme}
quantum PD
\index{Quantum Prisoners' Dilemma}
assigns two basis vectors $\left|  C\right\rangle $ and $\left|
D\right\rangle $ in the Hilbert space
\index{Hilbert space}
of a qubit
\index{Qubit}%
. States of the two qubits belong to two-dimensional Hilbert spaces
\index{Hilbert space}
$H_{A}$ and $H_{B}$, respectively. The state of the game is defied as being a
vector residing in the tensor-product space
\index{Tensor-product space}
$H_{A}\otimes H_{B}$, spanned by the basis $\left|  CC\right\rangle ,\left|
CD\right\rangle ,\left|  DC\right\rangle $ and $\left|  DD\right\rangle $.
Game's initial state is $\left|  \psi_{ini}\right\rangle =\hat{J}\left|
CC\right\rangle $ where $\hat{J}$ is a unitary operator
\index{Unitary operator}
known to both the players. Alice's and Bob's strategies are unitary operations
$\hat{U}_{A}$ and $\hat{U}_{B}$, respectively, chosen from a strategic space
\c{S}. After players' actions the state of the game changes to $\hat{U}%
_{A}\otimes\hat{U}_{B}\hat{J}\left|  CC\right\rangle $. Finally, the state is
measured and it consists of applying reverse unitary operator
\index{Unitary operator}
$\hat{J}^{\dagger}$ followed by a pair of Stern-Gerlach type detectors
\index{Stern-Gerlach detectors}%
. Before detection the final state of the game is $\left|  \psi_{fin}%
\right\rangle =\hat{J}^{\dagger}\hat{U}_{A}\otimes\hat{U}_{B}\hat{J}\left|
CC\right\rangle $. Players' expected payoffs are the projections of the state
$\left|  \psi_{fin}\right\rangle $ onto the basis vectors of tensor-product
space
\index{Tensor-product space}
$H_{A}\otimes H_{B}$, weighed by the constants appearing in the following game
matrix (\ref{PD matrix}).%

\begin{equation}%
\begin{array}
[c]{c}%
\text{Alice}%
\end{array}%
\begin{array}
[c]{c}%
C\\
D
\end{array}
\overset{\overset{%
\begin{array}
[c]{c}%
\text{Bob}%
\end{array}
}{%
\begin{array}
[c]{cc}%
C & D
\end{array}
}}{\left(
\begin{array}
[c]{cc}%
(r,r) & (s,t)\\
(t,s) & (u,u)
\end{array}
\right)  } \label{PD matrix}%
\end{equation}
where $C$ and $D$ are the classical strategies of Cooperation and Defection,
respectively. The first and the second entry in small braces correspond to
Alice's and Bob's (classical, pure strategy) payoffs, respectively. When
$s<u<r<t$ the matrix (\ref{PD matrix}) represents PD. In EWL quantum PD
\index{EWL quantization scheme}
Alice's payoff, for example, reads%

\begin{equation}
P_{A}=r\left|  \left\langle CC\mid\psi_{fin}\right\rangle \right|
^{2}+s\left|  \left\langle CD\mid\psi_{fin}\right\rangle \right|
^{2}+t\left|  \left\langle DC\mid\psi_{fin}\right\rangle \right|
^{2}+u\left|  \left\langle DD\mid\psi_{fin}\right\rangle \right|  ^{2}\text{.}
\label{Eisert's Alice's payoff}%
\end{equation}
With reference to the matrix (\ref{PD matrix}) Bob's payoff is, then, obtained
by the transformation $s\rightleftarrows t$ in Eq.
(\ref{Eisert's Alice's payoff}). Eisert and Wilkens \cite{Eisert1} used
following matrix representations of unitary operators
\index{Unitary operator}
of their one- and two-parameter
\index{One-parameter strategy set}
\index{Two-parameter strategy set}
strategies, respectively:%

\begin{align}
U(\theta)  &  =\left(
\begin{array}
[c]{cc}%
\cos(\theta/2) & \sin(\theta/2)\\
\text{-}\sin(\theta/2) & \cos(\theta/2)
\end{array}
\right) \label{OneParameterSet}\\
U(\theta,\phi)  &  =\left(
\begin{tabular}
[c]{ll}%
e$^{i\phi}\cos(\theta/2)$ & $\sin(\theta/2)$\\
$\text{-}\sin(\theta/2)$ & e$^{-i\phi}\cos(\theta/2)$%
\end{tabular}
\right)  \label{TwoParameterSet}%
\end{align}
where%

\begin{equation}
0\leq\theta\leq\pi\text{ and }0\leq\phi\leq\pi/2\text{.} \label{Ranges}%
\end{equation}
To ensure that the classical game is faithfully represented in its quantum
version, EWL imposed an additional conditions on $\hat{J}$:%

\begin{equation}
\left[  \hat{J},\hat{D}\otimes\hat{D}\right]  =0,\left[  \hat{J},\hat
{D}\otimes\hat{C}\right]  =0,\left[  \hat{J},\hat{C}\otimes\hat{D}\right]  =0
\label{condition1}%
\end{equation}
with $\hat{C}$ and $\hat{D}$ being the operators corresponding to the
classical strategies $C$ and $D$, respectively. A unitary operator
\index{Unitary operator}
satisfying the condition (\ref{condition1}) is%

\begin{equation}
\hat{J}=\exp\left\{  i\gamma\hat{D}\otimes\hat{D}/2\right\}
\end{equation}
where $\gamma\in\left[  0,\pi/2\right]  $ and $\hat{J}$ represents a measure
of the game's entanglement
\index{Entanglement}%
. At $\gamma=0$ the game can be interpreted as a mixed-strategy classical
game. For a maximally entangled game $\gamma=\pi/2$ the classical NE of
$\hat{D}\otimes\hat{D}$ is replaced by a different unique equilibrium $\hat
{Q}\otimes\hat{Q}$ where $\hat{Q}\sim\hat{U}(0,\pi/2).$ This new equilibrium
is found also to be \emph{Pareto optimal}
\index{Pareto optimal}%
\textit{\ }\cite{Rasmusen89}, that is, a player cannot increase his/her payoff
by deviating from this pair of strategies without reducing the other player's
payoff. Classically $(C,C)$ is Pareto optimal, but is not an equilibrium
\cite{Rasmusen89}, thus resulting in the `dilemma' in the game. It is argued
\cite{Benjamin1,Eisert's reply} that in its quantum version the dilemma
disappears from the game and quantum strategies
\index{Quantum strategy}
give a superior performance if entanglement
\index{Entanglement}
is present.%

\begin{figure}
[ptb]
\begin{center}
\includegraphics[
trim=0.000000in 0.000000in -0.042379in -0.008210in,
height=2.4327in,
width=3.749in
]%
{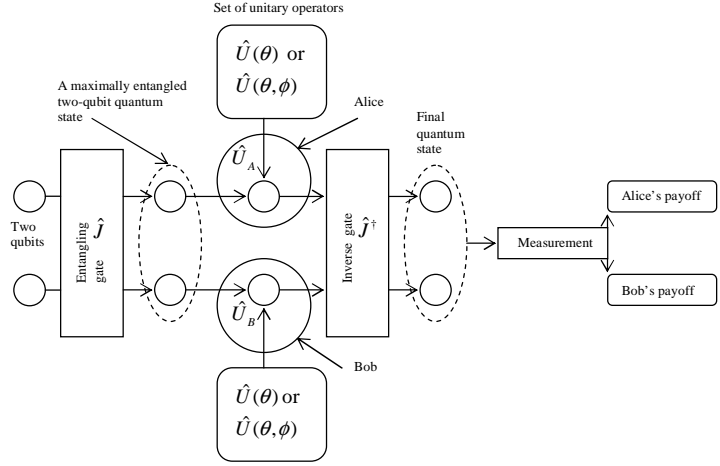}%
\caption{EWL scheme to play a quantum game.}%
\label{Fig1}%
\end{center}
\end{figure}

MW quantization scheme
\index{MW quantization scheme}
\cite{Marinatto1,Benjamin2,Marinatto's reply} for BoS
\index{Quantum Battle of Sexes}
\index{Battle of Sexes (BoS)}
identifies a state in $2\otimes2$ dimensional Hilbert space
\index{Hilbert space}
as a \emph{strategy}. At the start of the game the players are supplied with
this strategy and the players manipulate the strategy in the next phase by
playing their \emph{tactics}. The state is finally measured and payoffs are
rewarded depending on the results of the measurement. A player can do actions
within a two-dimensional subspace. Tactics
\index{Tactics}
are therefore \emph{local actions}
\index{Local actions!on a qubit}
on a player's qubit. The final measurement, made independently on each qubit
\index{Qubit}%
, takes into consideration the local nature of players' manipulations. This is
done by selecting a measurement basis that respects the division of Hilbert
space
\index{Hilbert space}
into two equal parts.

Essentially MW scheme
\index{MW quantization scheme}
differs from EWL scheme
\index{EWL quantization scheme}
\cite{Eisert,Eisert1} in the absence of reverse gate\footnote{EWL introduced
the gate $J^{\dagger}$ before measurement takes place that makes sure that the
classical game remains a subset of its quantum version.} $J^{\dagger}$.
Finally, the quantum state is measured and it is found that the classical game
remains a subset of the quantum game if the players' tactics are limited to a
convex linear combination, with real coefficients, of applying the identity
$\hat{I}$ and the Pauli spin-flip operator
\index{Pauli spin-flip operator}
$\hat{\sigma}_{x}$. Classical game results when the players are forwarded an
initial strategy $\left|  \psi_{in}\right\rangle =\left|  00\right\rangle $.%

\begin{figure}
[ptb]
\begin{center}
\includegraphics[
height=2.1456in,
width=3.6409in
]%
{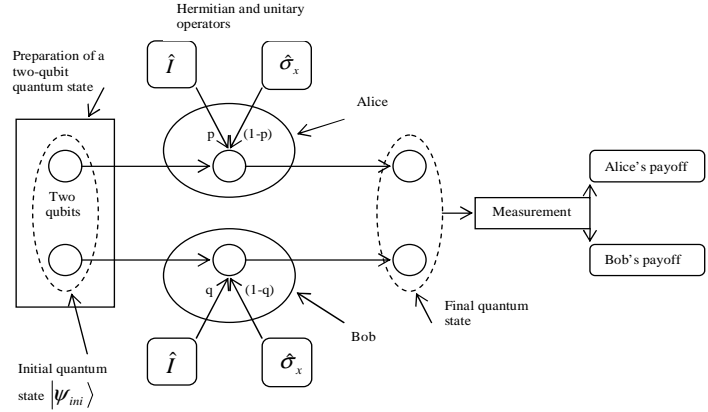}%
\caption{MW scheme to play a quantum game.}%
\label{Fig2}%
\end{center}
\end{figure}

Let $\rho_{in}$ be the initial strategy the players Alice and Bob receive at
the start of the game. Assume Alice acts with identity $\hat{I}$ on $\rho
_{in}$ with probability $p$ and with $\hat{\sigma}_{x}$ with probability
$(1-p)$. Similarly, Bob act with identity $\hat{I}$ with probability $q$ and
with $\hat{\sigma}_{x}$ with probability $(1-q)$. After players' actions the
state changes to%

\begin{align}
\rho_{fin}  &  =pq\hat{I}_{A}\otimes\hat{I}_{B}\rho_{in}\hat{I}_{A}^{\dagger
}\otimes\hat{I}_{B}^{\dagger}+p(1-q)\hat{I}_{A}\otimes\hat{\sigma}_{xB}%
\rho_{in}\hat{I}_{A}^{\dagger}\otimes\hat{\sigma}_{xB}^{\dagger}+\nonumber\\
&  q(1-p)\hat{\sigma}_{xA}\otimes\hat{I}_{B}\rho_{in}\hat{\sigma}%
_{xA}^{\dagger}\otimes\hat{I}_{B}^{\dagger}+\nonumber\\
&  (1-p)(1-q)\hat{\sigma}_{xA}\otimes\hat{\sigma}_{xB}\rho_{in}\hat{\sigma
}_{xA}^{\dagger}\otimes\hat{\sigma}_{xB}^{\dagger}\text{.}%
\end{align}
When the game is given by the bi-matrix:%

\begin{equation}%
\begin{array}
[c]{c}%
\text{Alice}%
\end{array}%
\begin{array}
[c]{c}%
S_{1}\\
S_{2}%
\end{array}
\overset{\overset{%
\begin{array}
[c]{c}%
\text{Bob}%
\end{array}
}{%
\begin{array}
[c]{cc}%
S_{1} & S_{2}%
\end{array}
}}{\left(
\begin{array}
[c]{cc}%
(\alpha_{A},\alpha_{B}) & (\beta_{A},\beta_{B})\\
(\gamma_{A},\gamma_{B}) & (\delta_{A},\delta_{B})
\end{array}
\right)  }%
\end{equation}
the payoff operators
\index{Payoff operators}
are:%

\begin{align}
(P_{A})_{oper}  &  =\alpha_{A}\left|  00\right\rangle \left\langle 00\right|
+\beta_{A}\left|  01\right\rangle \left\langle 01\right|  +\gamma_{A}\left|
10\right\rangle \left\langle 10\right|  +\delta_{A}\left|  11\right\rangle
\left\langle 11\right| \nonumber\\
(P_{B})_{oper}  &  =\alpha_{B}\left|  00\right\rangle \left\langle 00\right|
+\beta_{B}\left|  01\right\rangle \left\langle 01\right|  +\gamma_{B}\left|
10\right\rangle \left\langle 10\right|  +\delta_{B}\left|  11\right\rangle
\left\langle 11\right|  \text{.}%
\end{align}
and payoff functions are then obtained as mean values of these operators:%

\begin{equation}
P_{A,B}=Tr\left\{  (P_{A,B})_{oper}\rho_{fin}\right\}  \text{.}%
\end{equation}

It is to be pointed out that in EWL set-up a quantum game corresponds when the
entanglement parameter $\gamma$ of the initial quantum state is different from
zero. When $\gamma$ is non-zero the players have strategies available to them
that result in the classical game. The general idea is to allow a range of
values to the parameter $\gamma$ and then to find how it leads to a different,
i.e. non-classical, equilibrium in the game.

In MW scheme
\index{MW quantization scheme}
\cite{Marinatto1,Benjamin2,Marinatto's reply} an initial strategy is forwarded
to two players who then apply their tactics on it and the classical game
corresponds to the initial state $\left|  00\right\rangle $. Assume now that
the players receive pure two-qubit states, different from $\left|
00\right\rangle $, and the measurement remains the same. A quantum form of the
game then corresponds if initial states different from the product state
$\left|  00\right\rangle $ are used. This translates finding quantum form of a
matrix game to finding appropriate initial state(s). This is justified because
the only restriction \cite{Marinatto's reply} on a quantum form of a game
being that the corresponding classical game must be reproducible as its
special case. As the product initial state $\left|  00\right\rangle $ always
results in the classical game, this approach remains within the mentioned restriction.

In EWL scheme
\index{EWL quantization scheme}
one looks for new equilibria in games in relation to the parameter $\gamma$.
In the above approach, however, one finds equilibria in relation to different
initial states. In this chapter, we will restrict ourselves to pure states
\index{Pure states}
only.

\section{Evolutionary stability
\index{Evolutionary stability}
in quantum games
\index{Quantum games}%
}

The concept of a NE was addressed in the earliest research publications in
quantum games
\index{Quantum games}
\cite{Eisert, Eisert1}. Analysis of this solution concept from non-cooperative
game theory
\index{Non-cooperative game theory}
generated significant interest in the new research field. These publications
do not explicitly refer to the population interpretation of the NE concept. In
fact, this possibility of this interpretation was behind the development of
the ESS
\index{ESS}
concept in evolutionary game theory. And when this interpretation is brought
within the domain of quantum games
\index{Quantum games}
it becomes natural to consider ESSs in this domain.

One may ask how and where the population setting may be relevant to quantum
games
\index{Quantum games}%
. How can a setting, originally developed to model the population biology
problems, be relevant and useful to quantum games
\index{Quantum games}%
? One can often sharpen this argument given the fact that, to date, almost all
of the experimental realizations of quantum games
\index{Quantum games}
are artificially constructed in laboratories using quantum computational
circuits \cite{Nielsen}.

Several replies can be made to this question, for example, that this setting
was behind the development of the NE concept that was addressed in the
earliest constructions of quantum games
\index{Quantum games}
attracting significant attention. One also finds that evolutionary stability
\index{Evolutionary stability}
has very rich literature in game theory, mathematical biology
\index{Mathematical biology}
and in evolutionary economics
\index{Evolutionary economics}
\cite{Friedman,Evolutionary economics}. In quantum games
\index{Quantum games}
the NE has been discussed in relation to quantum entanglement
\index{Quantum entanglement}
\cite{Nielsen} and the possibility that the same can be done with evolutionary
stability
\index{Evolutionary stability}
clearly opens a new interesting role for this quantum phenomenon. It is
conjectured that the possibility of this extended role for entanglement may
perhaps be helpful to better understand entanglement itself.

Evolutionary stability
\index{Evolutionary stability}
presents a game-theoretic model to understand evolutionary dynamics. Recent
developments in quantum games
\index{Quantum games}
motivate to ask how this game-theoretic solution concept adapts/shapes/changes
itself when players are given access to quantum strategies
\index{Quantum strategy}%
. This questions is clearly related to a bigger question: Can quantum
mechanics
\index{Quantum mechanics}
have a role in directing, or possibly dictating, the dynamics of evolution? We
believe that for an analysis along this direction the evolutionary stability
\index{Evolutionary stability}
offers an interesting situation because, firstly, it is a simple and a
beautiful concept and, secondly, it is supported by extensive literature
\cite{Bomze,Hofbauer}.

To discuss evolutionary stability
\index{Evolutionary stability}
in quantum games
\index{Quantum games}
may appear as if a concept originally developed for population biology
problems is arbitrarily being placed within the domain of quantum games. One
can reply to this by noticing that population biology is not the only relevant
domain for the concept of evolutionary stability
\index{Evolutionary stability}%
. This concept can also be interpreted using infinitely repeated two-player
games and without referring to a population of players. Secondly, as the
Nash's thesis \cite{NashThesis,Hofbauer} showed it, it is not the population
biology alone that motivates a population setting for game theory --
responsible for the concept of evolutionary stability
\index{Evolutionary stability}%
. Surprisingly, the concept of NE also does the same, although it may not be
recognized generally.

The usual approach in game theory consists of analyzing games among
hyper-rational players who are always found both ready and engaged in their
selfish interests to optimize their payoffs or utilities. Evolutionary
stability
\index{Evolutionary stability}
has roots in the efforts to get rid of this usual approach that game theory
had followed. The lesson it teaches is that playing games can be disassociated
from players' capacity to make rational decisions. This disassociation seems
equally valid in those possible situations where nature plays quantum games
\footnote{Although, no evidence showing nature playing quantum games has been
found to date, the idea itself does not seem far-fetched.}. It is because
associating rationality to quantum-interacting entities is of even a more
remote possibility than it is the case when this association is made to
bacteria and viruses, whose behavior evolutionary game theory explains.

In the following we will try to address the following questions: How ESSs are
affected when a classical game, played by a population, changes itself to a
quantum form? How pure and mixed ESSs
\index{ESS}
are distinguished from one another when such a change in the form of a game
takes place? Can quantum games provide a route that can relate evolutionary
dynamics, for example, to quantum entanglement
\index{Quantum entanglement}%
? Considering a population of players in which a classical strategy has
established itself as an ESS
\index{ESS}%
, we would like to ask: a) What happens when `mutants' of ESS theory come up
with quantum strategies
\index{Quantum strategy}
and try to invade the classical ESS? b) What happens if such an invasion is
successful and a new ESS is established -- an ESS that is quantum in nature?
c) Suppose afterwards another small group of mutants appears which is equipped
with some other quantum strategy
\index{Quantum strategy}%
. Will it successfully invade the quantum ESS?

\subsection{Evolutionary stability
\index{Evolutionary stability}
in EWL scheme
\index{EWL quantization scheme}%
}

EWL used the matrix (\ref{PD matrix}) with $r=3,s=0,t=5,$ and $u=1$ in their
proposal for quantum PD. Assume a population setting where in each pair-wise
encounter the players play PD with the same matrix and each contest is
symmetric. Which strategies will then be likely to be stable? Straightforward
analysis \cite{Prestwich} shows that $D$ will be the pure classical strategy
prevalent in the population and hence the classical ESS. We consider following
three cases:

Case (a) A small group of mutants appear equipped with one-parameter quantum
strategy
\index{Quantum strategy}
$\hat{U}(\theta)$ when $D$ exists as a classical ESS

Case (b) Mutants are equipped with two-parameter quantum strategy
\index{Quantum strategy}
$\hat{U}(\theta,\phi)$ against the classical ESS

Case (c) Mutants have successfully invaded and a two-parameter quantum
strategy $\hat{Q}\sim\hat{U}(0,\pi/2)$ has established itself as a new quantum
ESS. Again another small group of mutants appear, using some other
two-parameter quantum strategy, and tries to invade the quantum ESS, which is
$\hat{Q}$.

\paragraph{Case (a):}

Because players are anonymous one can represent $P(\hat{U}(\theta),D)$ as the
payoff to $\hat{U}(\theta)$-player against the $D$-player. Here $\hat
{U}(\theta)$ is the Eisert and Wilkens' one-parameter quantum strategy set
(\ref{OneParameterSet}). Players' payoffs read $P(\hat{U}(\theta),D)=\sin
^{2}(\theta/2)$; $P(\hat{U}(\theta),\hat{U}(\theta))=2\cos^{2}(\theta
/2)+5\cos^{2}(\theta/2)\sin^{2}(\theta/2)+1$; $P(D,\hat{U}(\theta))=5\cos
^{2}(\theta/2)+\sin^{2}(\theta/2)$; and $P(D,D)=1$. Now $P(D,D)>P(\hat
{U}(\theta),D)$ for all $\theta\in\lbrack0,\pi)$. Hence the first condition
for an ESS holds and $D\sim\hat{U}(\pi)$ is an ESS. The case $\theta=\pi$
corresponds to one-parameter mutant strategy
\index{Mutant strategy}
coinciding with the ESS, which is ruled out. If $D\sim\hat{U}(\pi)$ is played
by almost all the members of the population -- which corresponds to high
frequency $F_{D}$ for $D$ -- we then have $W(D)>W(\theta)$ for all $\theta
\in\lbrack0,\pi)$ using the definition (\ref{fitnesses}). The fitness of a
one-parameter quantum strategy\footnote{In EWL set-up one-parameter quantum
strategies correspond to mixed (randomized) classical strategies.}, therefore,
cannot exceed the fitness of a classical ESS. And a one-parameter quantum
strategy cannot invade a classical ESS.

\paragraph{Case (b):}

Let $\hat{U}(\theta,\phi)$ be a two-parameter strategy from the set
(\ref{TwoParameterSet}). The expected payoffs read $P(D,D)=1$; $P(D,\hat
{U}(\theta,\phi))=5\cos^{2}(\phi)\cos^{2}(\theta/2)+\sin^{2}(\theta/2)$;
$P(\hat{U}(\theta,\phi),D)=5\sin^{2}(\phi)\cos^{2}(\theta/2)+\sin^{2}%
(\theta/2) $; and%

\begin{align}
P(\hat{U}(\theta,\phi),\hat{U}(\theta,\phi))  &  =3\left|  \cos(2\phi)\cos
^{2}(\theta/2)\right|  ^{2}\nonumber\\
&  +5\cos^{2}(\theta/2)\sin^{2}(\theta/2)\left|  \sin(\phi)-\cos(\phi)\right|
^{2}\nonumber\\
&  +\left|  \sin(2\phi)\cos^{2}(\theta/2)+\sin^{2}(\theta/2)\right|
^{2}\text{.}%
\end{align}
Here $P(D,D)>P(\hat{U}(\theta,\phi),D)$ if $\phi<\arcsin(1/\sqrt{5})$ and if
$P(D,D)=P(\hat{U}(\theta,\phi),D)$ then $P(D,\hat{U}(\theta,\phi))>P(\hat
{U}(\theta,\phi),\hat{U}(\theta,\phi))$. Therefore, $D$ is an ESS if
$\phi<\arcsin(1/\sqrt{5})$ otherwise the strategy $\hat{U}(\theta,\phi)$ will
be in position to invade $D$. Alternatively, if most of the members of the
population play $D\sim\hat{U}(\pi,0)$ -- which means a high frequency $F_{D}$
for $D$ -- then the fitness $W(D)$ will remain greater than the fitness
$W[\hat{U}(\theta,\phi)]$ if $\phi<\arcsin(1/\sqrt{5})$. For $\phi
>\arcsin(1/\sqrt{5})$ the strategy $\hat{U}(\theta,\phi)$ can invade the
strategy $D$, which is the classical ESS.

In this analysis mutants are able to invade $D$ when $\phi>\arcsin(1/\sqrt
{5})$ and the invasion may seem not so unusual given the fact that they
exploit richer strategies. But it leads to the third case i.e. when `quantum
mutants
\index{Quantum mutants}%
' have successfully invaded and a two-parameter strategy $\hat{U}$ has
established itself. Can now some new mutants coming up with $\hat{Q}\sim
\hat{U}(0,\pi/2)$ and invade the `quantum ESS'?

\paragraph{Case (c):}

EWL \cite{Eisert,Eisert1} showed that in their quantum PD the quantum strategy
$\hat{Q}$, played by both the players, is the unique NE. How mutants playing
$\hat{Q}$ come up against $\hat{U}(\theta,\phi)$ which already exists as an
ESS? To find it the following payoffs are obtained. $P(\hat{Q},\hat{Q})=3$;
$P(\hat{U}(\theta,\phi),\hat{Q})=[3-2\cos^{2}(\phi)]\cos^{2}(\theta/2)$; and
$P(\hat{Q},\hat{U}(\theta,\phi))=[3-2\cos^{2}(\phi)]\cos^{2}(\theta
/2)+5\sin^{2}(\theta/2)$. Now the inequality $P(\hat{Q},\hat{Q})>P(\hat
{U}(\theta,\phi),\hat{Q})$ holds for all $\theta\in\lbrack0,\pi]$ and $\phi
\in\lbrack0,\pi/2]$ except when $\theta=0$ and $\phi=\pi/2$, which is the case
when the mutant strategy
\index{Mutant strategy}
$\hat{U}(\theta,\phi)$ is the same as $\hat{Q}$. This case is obviously ruled
out. The first condition for $\hat{Q}$ to be an ESS, therefore, holds. The
condition $P(\hat{Q},\hat{Q})=P(\hat{U}(\theta,\phi),\hat{Q})$ implies
$\theta=0$ and $\phi=\pi/2$. Again we have the situation of mutant strategy
\index{Mutant strategy}
same as $\hat{Q}$ and the case is neglected. If $\hat{Q}$ is played by most of
the players, meaning high frequency $F_{\hat{Q}}$ for $\hat{Q}$, then
$W(\hat{Q})>W[\hat{U}(\theta,\phi)]$ for all $\theta\in(0,\pi]$ and $\phi
\in\lbrack0,\pi/2)$. A two-parameter quantum strategy $\hat{U}(\theta,\phi)$,
therefore, cannot invade the quantum ESS (i.e. the strategy $\hat{Q}\sim
\hat{U}(0,\pi/2)$). Mutants' access to richer strategies, as it happens in the
case (B), does not continue to be an advantage as most of the population also
have access to it. Hence $\hat{Q}$ comes out as the unique NE and ESS of the game.

\subsubsection{Evolutionary stability
\index{Evolutionary stability}
and entanglement}

Above analysis motivates to obtain a direct relationship between a measure of
entanglement
\index{Measure of entanglement}
and the mathematical concept of evolutionary stability
\index{Evolutionary stability}
for two-player games. The following example shows this relationship. Consider
the two-player game given by the matrix (\ref{General Matrix}):%

\begin{equation}%
\begin{array}
[c]{c}%
\text{Alice}%
\end{array}%
\begin{array}
[c]{c}%
S_{1}\\
S_{2}%
\end{array}
\overset{\overset{%
\begin{array}
[c]{c}%
\text{Bob}%
\end{array}
}{%
\begin{array}
[c]{cc}%
S_{1} & S_{2}%
\end{array}
}}{\left(
\begin{array}
[c]{cc}%
(r,r) & (s,t)\\
(t,s) & (u,u)
\end{array}
\right)  } \label{General Matrix}%
\end{equation}
and suppose Alice and Bob play the strategy $S_{1}$ with probabilities $p$ and
$q$, respectively. The strategy $S_{2}$ is then played with probabilities
$(1-p)$ and $(1-q)$ by Alice and Bob, respectively. We denote Alice's payoff
by $P_{A}(p,q)$ when she plays $p$ and Bob plays $q$. That is, Alice's and
Bob's strategies are now identified by the numbers $p,q\in\lbrack0,1]$,
without referring to $S_{1}$ and $S_{2}$. For the matrix (\ref{General Matrix}%
) Alice's payoff $P_{A}(p,q)$, for example, reads%

\begin{equation}
P_{A}(p,q)=rpq+sp(1-q)+t(1-p)q+u(1-p)(1-q)\text{.} \label{Payoffs}%
\end{equation}
Similarly, Bob's payoff $P_{B}(p,q)$ can be written. In this symmetric game we
have $P_{A}(p,q)=P_{B}(q,p)$ and, without using subscripts, $P(p,q)$, for
example, describes the payoff to $p$-player against $q$-player. In this game
the inequality%

\begin{equation}
P(p^{\ast},p^{\ast})-P(p,p^{\ast})\geqslant0 \label{NE}%
\end{equation}
says that the strategy $p^{\ast}$, played by both the players, is a NE. We
consider the case when%

\begin{equation}
s=t,\text{ \ \ }r=u\text{ and \ \ }(r-t)>0 \label{Further Constraint}%
\end{equation}
in the matrix (\ref{General Matrix}). In this case the inequality (\ref{NE})
along with the definition (\ref{Payoffs}) gives%

\begin{equation}
P(p^{\ast},p^{\ast})-P(p,p^{\ast})=(p^{\ast}-p)(r-t)(2p^{\ast}-1)
\label{PayoffDiff}%
\end{equation}
and the strategy $p^{\ast}=1/2$ comes out as a mixed NE. From the ESS
definition (\ref{ESSDef}) we get $P(1/2,1/2)-P(p,1/2)=0$ and the part a) of
the definition does not apply. Part b) of the definition (\ref{ESSDef}), then, gives%

\begin{equation}
P(1/2,p)-P(p,p)=(r-t)\left\{  2p(1-p)-1/2\right\}  \label{PayoffDiff1}%
\end{equation}
which can not be strictly greater than zero given $(r-t)>0$. For example, at
$p=0$ it becomes a negative quantity. Therefore, for the matrix game defined
by (\ref{General Matrix}) and (\ref{Further Constraint}) the strategy
$p^{\ast}=1/2$ is a symmetric NE, but it is not evolutionarily stable. Also,
at this equilibrium both players get $(r+t)/2$ as their payoffs.

Now consider the same game, defined by (\ref{General Matrix}) and
(\ref{Further Constraint}), when it is played by the set-up proposed by EWL.
We set $s_{A}\equiv(\theta_{A},\phi_{A})$ and $s_{B}\equiv(\theta_{B},\phi
_{B})$ to denote Alice's and Bob's strategies, respectively. Because the
quantum game is symmetric i.e. $P_{A}(s_{A},s_{B})=P_{B}(s_{B},s_{A})$ we can
write, as before, $P(s_{A},s_{B})$ for the payoff to $s_{A}$-player against
$s_{B}$-player. For the quantum form of the game defined by
(\ref{General Matrix},\ref{Further Constraint}) one finds%

\begin{equation}
P(s_{A},s_{B})=(1/2)(r-t)\left\{  1+\cos\theta_{A}\cos\theta_{B}+\sin
\theta_{A}\sin\theta_{B}\sin\gamma\sin(\phi_{A}+\phi_{B})\right\}  +t\text{.}
\label{PayoffsQG}%
\end{equation}
The definition of a NE gives $P(s^{\ast},s^{\ast})-P(s,s^{\ast})\geqslant0$
where $s=(\theta,\phi)$ and $s^{\ast}=(\theta^{\ast},\phi^{\ast})$. This
definition can be written as%

\begin{equation}
\left\{  \partial_{\theta}P\mid_{\theta^{\ast},\phi^{\ast}}(\theta^{\ast
}-\theta)+\partial_{\phi}P\mid_{\theta^{\ast},\phi^{\ast}}(\phi^{\ast}%
-\phi)\right\}  \geq0\text{.}%
\end{equation}
We search for a quantum strategy $s^{\ast}=(\theta^{\ast},\phi^{\ast})$ for
which both $\partial_{\theta}P\mid_{\theta^{\ast},\phi^{\ast}}$ and
$\partial_{\phi}P\mid_{\theta^{\ast},\phi^{\ast}}$ vanish at $\gamma=0$ and
which, at some other value of $\gamma$, is not zero. For the payoffs
(\ref{PayoffsQG}) the strategy $s^{\ast}=(\pi/2,\pi/4)$ satisfies these
conditions. For this strategy Eq. (\ref{PayoffsQG}) gives%

\begin{equation}
P(s^{\ast},s^{\ast})-P(s,s^{\ast})=(1/2)(r-t)\sin\gamma\left\{  1-\sin
(\phi+\pi/4)\sin\theta\right\}  \text{.} \label{NEQG}%
\end{equation}
At $\gamma=0$ the strategy $s^{\ast}=(\pi/2,\pi/4)$, when played by both the
players, is a NE and it rewards the players same as does the strategy
$p^{\ast}=1/2$ in the classical version of the game i.e. $(r+t)/2$. Also, then
we have $P(s^{\ast},s^{\ast})-P(s,s^{\ast})=0$ from Eq. (\ref{NEQG}) and the
ESS's second condition in (\ref{ESSDef}) applies. Use Eq. (\ref{PayoffsQG}) to evaluate%

\begin{align}
P(s^{\ast},s)-P(s,s)  &  =-(r-t)\cos^{2}(\theta)+\nonumber\\
&  (1/2)(r-t)\sin\gamma\sin\theta\left\{  \sin(\phi+\pi/4)-\sin\theta
\sin(2\phi)\right\}  \label{ESSSecondQ}%
\end{align}
which at $\gamma=0$ reduces to $P(s^{\ast},s)-P(s,s)=-(r-t)\cos^{2}(\theta)$,
that can assume negative values. The game's definition
(\ref{Further Constraint}) and the ESS's second condition in (\ref{ESSDef})
show that the strategy $s^{\ast}=(\pi/2,\pi/4)$ is not evolutionarily stable
at $\gamma=0$.

Now consider the case when $\gamma\neq0$ in order to know about the
evolutionary stability
\index{Evolutionary stability}
of the \emph{same} quantum strategy. From (\ref{Ranges}) we have both
$\sin\theta,\sin(\phi+\pi/4)\in\lbrack0,1]$ and Eq. (\ref{NEQG}) indicates
that $s^{\ast}=(\pi/2,\pi/4)$ remains a NE for all $\gamma\in\lbrack0,\pi/2]$.
The product $\sin(\phi+\pi/4)\sin\theta$ attains a value of $1$ only at
$s^{\ast}=(\pi/2,\pi/4)$ and remains less than $1$ otherwise. Eq. (\ref{NEQG})
shows that for $\gamma\neq0$ the strategy $s^{\ast}=(\pi/2,\pi/4)$ becomes a
strict NE for which the ESS's first condition in (\ref{ESSDef}) applies.
Therefore, for the game defined in (\ref{Further Constraint}) the strategy
$s^{\ast}=(\pi/2,\pi/4)$ is evolutionarily stable for a non-zero measure of
entanglement
\index{Measure of entanglement}
$\gamma$. That is, entanglement gives evolutionary stability to a symmetric NE
by making it a strict NE, that is, it is achieved by using in (\ref{ESSDef})
the ESS's first condition only. Perhaps, a more interesting example would be
the case when entanglement gives evolutionary stability via the ESS's second
condition. In that case, entanglement will make $P(s^{\ast},s)$ strictly
greater than $P(s,s)$ when $P(s^{\ast},s^{\ast})$ and $P(s,s^{\ast})$ are equal.

It is to be pointed out here that in literature there exists an approach
\cite{Entropy} which characterizes ESSs in terms of extremal states of a
function known as \textit{evolutionary entropy}
\index{Evolutionary entropy}
that is defined by%

\begin{equation}
E=-\underset{i}{\sum}\mu_{i}\log\mu_{i}%
\end{equation}
where $\mu_{i}$ represents the relative contribution of the $i$-th strategy to
the total payoff. A possible extension of the present approach may be the case
when quantum entanglement
\index{Quantum entanglement}
decides extremal states of evolutionary entropy. Extension along similar lines
can be proposed for another quantity called \textit{relative negentropy}
\index{Relative negentropy}
\cite{Bomze} that is optimized during the course of evolution.

\subsection{Evolutionary stability in MW quantization scheme
\index{MW quantization scheme}%
}

Another interesting route that allows to consider evolutionary stability in
relation to quantization of a game
\index{Quantization of a game}
is provided by MW scheme
\index{MW quantization scheme}
\cite{Marinatto1}. In this scheme a transition between classical and quantum
game is achieved by the initial state: classical payoffs are obtained when the
initial state is a product state $\left|  \psi_{in}\right\rangle =\left|
00\right\rangle $. In this scheme one can consider evolutionary stability in a
quantum game by asking whether it possible that a particular symmetric NE
switches-over between being an ESS and not being an ESS when the initial state
(initial strategy) changes from being $\left|  \psi_{in}\right\rangle =\left|
00\right\rangle $ to another state. MW scheme
\index{MW quantization scheme}
offers the possibility to make transition from classical to quantum version of
a game by using different initial states and it appears to be a more suitable
quantization scheme to analyze evolutionary stability in quantum games. It is because:

a) In a symmetric bi-matrix game
\index{Bi-matrix game}%
, played in a population setting, players have access to two pure strategies
and a mixed strategy is interpreted as a convex linear combination of pure
strategies. Similar is the case with the players' strategies in MW scheme
\index{MW quantization scheme}
where a mixed strategy consists of a convex linear combination of the players'
actions with two unitary operators
\index{Unitary operator}%
.

b) Fitness of a pure strategy\ can be given a straightforward extension in MW
scheme
\index{MW quantization scheme}%
. It corresponds to a situation when, for example, in the quantum game, a
player uses only one unitary operator
\index{Unitary operator}
out of the two.

c) Theory of ESSs, in the classical domain, deals with anonymous players
possessing discrete number of pure strategies. EWL scheme
\index{EWL quantization scheme}
involves a continuum of pure quantum strategies
\index{Quantum strategy}%
. The ESS concept is known to encounter problems \cite{Oechssler} when players
possess a continuum of pure strategies.

\subsubsection{$2\times2$ asymmetric games}

An ESS is defined as a strict NE \cite{Weibull} for an asymmetric bi-matrix
game, i.e. the game $G=(M,N)$ for which $N\neq M^{T}$. That is, a strategy
pair $(\overset{\star}{x},\overset{\star}{y})\in S$ is an ESS of the game $G$
if $P_{A}(\overset{\star}{x},\overset{\star}{y})>P_{A}(x,\overset{\star}{y})$
and $P_{B}(\overset{\star}{x},\overset{\star}{y})>P_{B}(\overset{\star}{x},y)$
for all $x\neq\overset{\star}{x}$ and $y\neq\overset{\star}{y}$. For example,
the BoS:%

\begin{equation}
\left(
\begin{array}
[c]{cc}%
(\alpha,\beta) & (\gamma,\gamma)\\
(\gamma,\gamma) & (\beta,\alpha)
\end{array}
\right)  \label{AsymmetricGame}%
\end{equation}
where $\alpha>\beta>\gamma$ is a asymmetric game with three classical NE
\cite{Marinatto1} given as 1) $\overset{\star}{p_{1}}=\overset{\star}{q_{1}%
}=0$ 2) $\overset{\star}{p_{2}}=\overset{\star}{q_{2}}=1$ and 3)
$\overset{\star}{p_{3}}=\frac{\alpha-\gamma}{\alpha+\beta-2\gamma}%
,\overset{\star}{q_{3}}=\frac{\beta-\gamma}{\alpha+\beta-2\gamma}$. Here the
NE 1) and 2) are also ESS's but 3) is not because of not being a strict NE.
When the asymmetric game (\ref{AsymmetricGame}) is played with the initial
state $\left|  \psi_{in}\right\rangle =a\left|  S_{1}S_{1}\right\rangle
+b\left|  S_{2}S_{2}\right\rangle $, where $S_{1}$ and $S_{2}$ are players'
pure classical strategies, the following three NE \cite{Marinatto1} emerge 1)
$\overset{\star}{p_{1}}=\overset{\star}{q_{1}}=1$ 2) $\overset{\star}{p_{2}%
}=\overset{\star}{q_{2}}=0$ and 3) $\overset{\star}{p_{3}}=\frac
{(\alpha-\gamma)\left|  a\right|  ^{2}+(\beta-\gamma)\left|  b\right|  ^{2}%
}{\alpha+\beta-2\gamma},\overset{\star}{q_{3}}=\frac{(\alpha-\gamma)\left|
b\right|  ^{2}+(\beta-\gamma)\left|  a\right|  ^{2}}{\alpha+\beta-2\gamma}$.
It turns out that, similar to the classical case, the quantum NE $1$) and $2$)
are ESSs while $3$) is not. Now, play thsi game with a different initial state:%

\begin{equation}
\left|  \psi_{in}\right\rangle =a\left|  S_{1}S_{2}\right\rangle +b\left|
S_{2}S_{1}\right\rangle \label{antisymmetricState}%
\end{equation}
for which players' payoffs are:%

\begin{align}
P_{A}(p,q)  &  =p\left\{  -q(\alpha+\beta-2\gamma)+\alpha\left|  a\right|
^{2}+\beta\left|  b\right|  ^{2}-\gamma\right\}  +\nonumber\\
&  q\left\{  \alpha\left|  b\right|  ^{2}+\beta\left|  a\right|  ^{2}%
-\gamma\right\}  +\gamma\nonumber\\
P_{B}(p,q)  &  =q\left\{  -p(\alpha+\beta-2\gamma)+\beta\left|  a\right|
^{2}+\alpha\left|  b\right|  ^{2}-\gamma\right\}  +\nonumber\\
&  p\left\{  \beta\left|  b\right|  ^{2}+\alpha\left|  a\right|  ^{2}%
-\gamma\right\}  +\gamma
\end{align}
and there is only one NE i.e. $\overset{\star}{p}=\frac{\beta\left|  a\right|
^{2}+\alpha\left|  b\right|  ^{2}-\gamma}{\alpha+\beta-\gamma},\overset{\star
}{q_{3}}=\frac{\alpha\left|  a\right|  ^{2}+\beta\left|  b\right|  ^{2}%
-\gamma}{\alpha+\beta-\gamma}$, which is not an ESS. So that, no ESS exists
when BoS is played with the state (\ref{antisymmetricState}).

Consider now another game:%

\begin{equation}
\left(
\begin{array}
[c]{cc}%
(\alpha_{1},\alpha_{2}) & (\beta_{1},\beta_{2})\\
(\gamma_{1},\gamma_{2}) & (\sigma_{1},\sigma_{2})
\end{array}
\right)
\end{equation}
for which%

\begin{equation}
\left(
\begin{array}
[c]{cc}%
\alpha_{1} & \beta_{1}\\
\gamma_{1} & \sigma_{1}%
\end{array}
\right)  \neq\left(
\begin{array}
[c]{cc}%
\alpha_{2} & \beta_{2}\\
\gamma_{2} & \sigma_{2}%
\end{array}
\right)  ^{T}%
\end{equation}
and that it is played by using initial state $\left|  \psi_{in}\right\rangle
=a\left|  S_{1}S_{1}\right\rangle +b\left|  S_{2}S_{2}\right\rangle $ with
$\left|  a\right|  ^{2}+\left|  b\right|  ^{2}=1$. Players' payoffs are:%

\begin{align}
P_{A,B}(p,q)  &  =\alpha_{1,2}\left\{  pq\left|  a\right|  ^{2}%
+(1-p)(1-q)\left|  b\right|  ^{2}\right\} \nonumber\\
&  +\beta_{1,2}\left\{  p(1-q)\left|  a\right|  ^{2}+q(1-p)\left|  b\right|
^{2}\right\} \nonumber\\
&  +\gamma_{1,2}\left\{  p(1-q)\left|  b\right|  ^{2}+q(1-p)\left|  a\right|
^{2}\right\} \nonumber\\
&  +\sigma_{1,2}\left\{  pq\left|  b\right|  ^{2}+(1-p)(1-q)\left|  a\right|
^{2}\right\}  \text{.}%
\end{align}
The NE conditions are%

\begin{gather}
P_{A}(\overset{\star}{p},\overset{\star}{q})-P_{A}(p,\overset{\star}%
{q})=\nonumber\\
(\overset{\star}{p}-p)\left[  \left|  a\right|  ^{2}(\beta_{1}-\sigma
_{1})+\left|  b\right|  ^{2}(\gamma_{1}-\alpha_{1})-\overset{\star}{q}\left\{
(\beta_{1}-\sigma_{1})+(\gamma_{1}-\alpha_{1})\right\}  \right]  \geq0\\
P_{B}(\overset{\star}{p},\overset{\star}{q})-P_{B}(\overset{\star}%
{p},q)=\nonumber\\
(\overset{\star}{q}-q)\left[  \left|  a\right|  ^{2}(\gamma_{2}-\sigma
_{2})+\left|  b\right|  ^{2}(\beta_{2}-\alpha_{2})-\overset{\star}{p}\left\{
(\gamma_{2}-\sigma_{2})+(\beta_{2}-\alpha_{2})\right\}  \right]  \geq0\text{.}%
\end{gather}
So that, for $\overset{\star}{p}=\overset{\star}{q}=0$ to be a NE we have%

\begin{align}
P_{A}(0,0)-P_{A}(p,0)  &  =-p\left[  (\beta_{1}-\sigma_{1})+\left|  b\right|
^{2}\left\{  (\gamma_{1}-\alpha_{1})-(\beta_{1}-\sigma_{1})\right\}  \right]
\geq0\nonumber\\
P_{B}(0,0)-P_{B}(0,q)  &  =-q\left[  (\gamma_{2}-\sigma_{2})+\left|  b\right|
^{2}\left\{  (\beta_{2}-\alpha_{2})-(\gamma_{2}-\sigma_{2})\right\}  \right]
\geq0\nonumber\\
&
\end{align}
and for the strategy pair $(0,0)$ to be an ESS in the classical
game\footnote{which corresponds when $\left|  b\right|  ^{2}=0$} we require
$P_{A}(0,0)-P_{A}(p,0)=-p(\beta_{1}-\sigma_{1})>0$ and $P_{B}(0,0)-P_{B}%
(0,q)=-q(\gamma_{2}-\sigma_{2})>0$ for all $p,q\neq0$. That is, $(\beta
_{1}-\sigma_{1})<0$ and $(\gamma_{2}-\sigma_{2})<0$. For the pair $(0,0)$ not
to be an ESS for some $\left|  b\right|  ^{2}\neq0$, let take $\gamma
_{1}=\alpha_{1\text{ }}$and $\beta_{2}=\alpha_{2}$ and we have%

\begin{align}
P_{A}(0,0)-P_{A}(p,0)  &  =-p(\beta_{1}-\sigma_{1})\left\{  1-\left|
b\right|  ^{2}\right\} \nonumber\\
P_{B}(0,0)-P_{B}(0,q)  &  =-q(\gamma_{2}-\sigma_{2})\left\{  1-\left|
b\right|  ^{2}\right\}
\end{align}
i.e. the pair $(0,0)$ doesn't remain an ESS at $\left|  b\right|  ^{2}=1$. A
game having this property is given by the matrix:%

\begin{equation}
\left(
\begin{array}
[c]{cc}%
(1,1) & (1,2)\\
(2,1) & (3,2)
\end{array}
\right)  \text{.}%
\end{equation}
For this game the strategy pair $(0,0)$ is an ESS when $\left|  b\right|
^{2}=0$ (classical game) but it is not when for example $\left|  b\right|
^{2}=\frac{1}{2}$, though it remains a NE in both the cases. The example shows
a NE switches between ESS and `not ESS' by using different initial state. In
contrast to the last case, one can also find initial states -- different from
the one corresponding to the classical game -- that turn a NE strategy pair
into an ESS. An example of a game for which it happens is%

\begin{equation}%
\begin{array}
[c]{c}%
\text{Alice}%
\end{array}%
\begin{array}
[c]{c}%
S_{1}\\
S_{2}%
\end{array}
\overset{\overset{%
\begin{array}
[c]{c}%
\text{Bob}%
\end{array}
}{%
\begin{array}
[c]{cc}%
S_{1} & S_{2}%
\end{array}
}}{\left(
\begin{array}
[c]{cc}%
(2,1) & (1,0)\\
(1,0) & (1,0)
\end{array}
\right)  }\text{.} \label{ExampleGame1}%
\end{equation}
Playing this game again via the state $\left|  \psi_{in}\right\rangle
=a\left|  S_{1}S_{1}\right\rangle +b\left|  S_{2}S_{2}\right\rangle $ gives
the following payoff differences for the strategy pair $(0,0)$:%

\begin{equation}
P_{A}(0,0)-P_{A}(p,0)=p\left|  b\right|  ^{2}\ \ \text{and}\ \ P_{B}%
(0,0)-P_{B}(0,q)=q\left|  b\right|  ^{2}%
\end{equation}
for Alice and Bob respectively. Therefore, (\ref{ExampleGame1}) is an example
of a game for which the pair $(0,0)$ is not an ESS when the initial state
corresponds to the classical game. But the same pair is an ESS for other
initial states for which $0<\left|  b\right|  ^{2}<1$.

\subsubsection{$2\times2$ symmetric games}

Consider now a symmetric bi-matrix game:%

\begin{equation}%
\begin{array}
[c]{c}%
\text{Alice}%
\end{array}%
\begin{array}
[c]{c}%
S_{1}\\
S_{2}%
\end{array}
\overset{\overset{%
\begin{array}
[c]{c}%
\text{Bob}%
\end{array}
}{%
\begin{array}
[c]{cc}%
S_{1} & S_{2}%
\end{array}
}}{\left(
\begin{array}
[c]{cc}%
(\alpha,\alpha) & (\beta,\gamma)\\
(\gamma,\beta) & (\delta,\delta)
\end{array}
\right)  } \label{PayoffMatrixGen2Player}%
\end{equation}
that is played by an initial state:%

\begin{equation}
\left|  \psi_{in}\right\rangle =a\left|  S_{1}S_{1}\right\rangle +b\left|
S_{2}S_{2}\right\rangle \text{, \ \ with }\left|  a\right|  ^{2}+\left|
b\right|  ^{2}=1\text{.} \label{IniStatGen2Player}%
\end{equation}
Let Alice's strategy consists of applying the identity operator $\hat{I}$ with
probability $p$ and the operator $\hat{\sigma}_{x}$ with probability $(1-p)$,
on the initial state written $\rho_{in}$ in density matrix
\index{Density matrix}
notation. Similarly Bob applies the operators $\hat{I}$ and $\hat{\sigma}_{x}$
with the probabilities $q$ and $(1-q)$ respectively. The final state is%

\begin{equation}
\rho_{fin}=\underset{\hat{U}=\hat{I},\hat{\sigma}_{x}}{\sum}\Pr(\hat{U}%
_{A})\Pr(\hat{U}_{B})[\hat{U}_{A}\otimes\hat{U}_{B}\rho_{in}\hat{U}%
_{A}^{\dagger}\otimes\hat{U}_{B}^{\dagger}]
\end{equation}
where unitary and Hermitian operator
\index{Hermitian operator}
$\hat{U}$ is either $\hat{I}$ or $\hat{\sigma}_{x}$. $\Pr(\hat{U}_{A})$,
$\Pr(\hat{U}_{B})$ are the probabilities, for Alice and Bob, respectively, to
apply the operator on the initial state. The matrix $\rho_{fin}$ is obtained
from $\rho_{in}$ by making a convex linear combination of players' possible
quantum operations. Payoff operators
\index{Payoff operators}
for Alice and Bob are \cite{Marinatto1}%

\begin{align}
(P_{A,B})_{oper}  &  =\alpha,\alpha\left|  S_{1}S_{1}\right\rangle
\left\langle S_{1}S_{1}\right|  +\beta,\gamma\left|  S_{1}S_{2}\right\rangle
\left\langle S_{1}S_{2}\right|  +\nonumber\\
&  \gamma,\beta\left|  S_{2}S_{1}\right\rangle \left\langle S_{2}S_{1}\right|
+\delta,\delta\left|  S_{2}S_{2}\right\rangle \left\langle S_{2}S_{2}\right|
\text{.}%
\end{align}
The payoffs are then obtained as mean values of these operators i.e.
$P_{A,B}=Tr\left[  (P_{A,B})_{oper}\rho_{fin}\right]  $. Because the quantum
game is symmetric with the initial state (\ref{IniStatGen2Player}) and the
payoff matrix
\index{Payoff matrix}
(\ref{PayoffMatrixGen2Player}), there is no need for subscripts. We can ,
then, write the payoff to a $p$-player against a $q$-player as $P(p,q)$, where
the first number is the focal player's move. When $\overset{\star}{p}$ is a NE
we find the following payoff difference:%

\begin{gather}
P(\overset{\star}{p},\overset{\star}{p})-P(p,\overset{\star}{p})=(\overset
{\star}{p}-p){\LARGE [}\left|  a\right|  ^{2}(\beta-\delta)+\nonumber\\
\left|  b\right|  ^{2}(\gamma-\alpha)-\overset{\star}{p}\left\{  (\beta
-\delta)+(\gamma-\alpha)\right\}  {\LARGE ]}\text{.}
\label{General2PlayerDifference1}%
\end{gather}
Now the ESS conditions for the pure strategy $p=0$ are given as%

\begin{gather}
1.\text{ \ \ \ }\left|  b\right|  ^{2}\left\{  (\beta-\delta)-(\gamma
-\alpha)\right\}  >(\beta-\delta)\nonumber\\
2.\text{ If }\left|  b\right|  ^{2}\left\{  (\beta-\delta)-(\gamma
-\alpha)\right\}  =(\beta-\delta)\nonumber\\
\text{then }q^{2}\left\{  (\beta-\delta)+(\gamma-\alpha)\right\}  >0
\end{gather}
where $1$ is the NE condition. Similarly the ESS conditions for the pure
strategy $p=1$ are%

\begin{gather}
1.\text{ \ \ \ }\left|  b\right|  ^{2}\left\{  (\gamma-\alpha)-(\beta
-\delta)\right\}  >(\gamma-\alpha)\nonumber\\
2.\text{ If }\left|  b\right|  ^{2}\left\{  (\gamma-\alpha)-(\beta
-\delta)\right\}  =(\gamma-\alpha)\nonumber\\
\text{then }(1-q)^{2}\left\{  (\beta-\delta)+(\gamma-\alpha)\right\}
>0\text{.}%
\end{gather}
Because these conditions, for both the pure strategies $p=1$ and $p=0$, depend
on $\left|  b\right|  ^{2}$, therefore, there can be examples of two-player
symmetric games for which the evolutionary stability of pure strategies can be
changed while playing the game using initial state in the form $\left|
\psi_{in}\right\rangle =a\left|  S_{1}S_{1}\right\rangle +b\left|  S_{2}%
S_{2}\right\rangle $. However, for the mixed NE, given as $\overset{\star}%
{p}=\frac{\left|  a\right|  ^{2}(\beta-\delta)+\left|  b\right|  ^{2}%
(\gamma-\alpha)}{(\beta-\delta)+(\gamma-\alpha)}$, the corresponding payoff
difference (\ref{General2PlayerDifference1}) becomes identically zero. From
the second condition of an ESS we find for the mixed NE $\overset{\star}{p}$
the difference%

\begin{align}
&  P(\overset{\star}{p},q)-P(q,q)=\frac{1}{(\beta-\delta)+(\gamma-\alpha
)}\times\nonumber\\
&  {\LARGE [}(\beta-\delta)-q\left\{  (\beta-\delta)+(\gamma-\alpha)\right\}
-\left|  b\right|  ^{2}\left\{  (\beta-\delta)-(\gamma-\alpha)\right\}
{\LARGE ]}^{2}\text{.}%
\end{align}
Therefore, the mixed strategy $\overset{\star}{p}$ is an ESS when $\left\{
(\beta-\delta)+(\gamma-\alpha)\right\}  >0$. This condition, making the mixed
NE $\overset{\star}{p}$ an ESS, is independent \footnote{An alternative
possibility is to adjust $\left|  b\right|  ^{2}$=$\frac{(\beta-\delta
)-q\left\{  (\beta-\delta)+(\gamma-\alpha)\right\}  }{\left\{  (\beta
-\delta)-(\gamma-\alpha)\right\}  }$ which makes the difference $\left\{
P(\overset{\star}{p},q)-P(q,q)\right\}  $ identically zero. The mixed strategy
$\overset{\star}{p}$ then does not remain an ESS. However such `mutant
dependent' adjustment of $\left|  b\right|  ^{2}$ is not reasonable because
the mutant strategy $q$ can be anything in the range $[0,1]$.} of $\left|
b\right|  ^{2}$. So that, in this symmetric two-player quantum game,
evolutionary stability of the mixed NE $\overset{\star}{p}$ can not be changed
when the game is played using initial quantum states of the form
(\ref{IniStatGen2Player}).

However, evolutionary stability of pure strategies can be affected, with this
form of the initial states, for two-player symmetric games. Examples of the
games with this property are easy to find. The class of games for which
$\gamma=\alpha$ and $(\beta-\delta)<0$ the strategies $p=0$ and $p=1$ remain
NE for all $\left|  b\right|  ^{2}\in\lbrack0,1]$; but the strategy $p=1$ is
not an ESS when $\left|  b\right|  ^{2}=0$ and the strategy $p=0$ is not an
ESS when $\left|  b\right|  ^{2}=1$.

Consider the symmetric bi-matrix game (\ref{PayoffMatrixGen2Player}) with the
constants $\alpha,\beta,\gamma,\delta$ satisfying the conditions:%

\begin{equation}
\alpha,\beta,\gamma,\delta\geq0;(\delta-\beta)>0;(\gamma-\alpha)\geq
0;(\gamma-\alpha)<(\delta-\beta)\text{.} \label{GameDefinition}%
\end{equation}
The condition making $(p^{\star},p^{\star})$ a NE is given by
(\ref{General2PlayerDifference1}). For this game three Nash equilibria arise
i.e. two pure strategies $p^{\ast}=0$, $p^{\ast}=1$, and one mixed strategy
$p^{\ast}=\frac{(\delta-\beta)\left|  a\right|  ^{2}-(\gamma-\alpha)\left|
b\right|  ^{2}}{(\delta-\beta)-(\gamma-\alpha)}$. These three cases are
considered below.

\paragraph{Case $p^{\star}=0:$}

For the strategy $p^{\star}=0$ to be a NE one requires%

\begin{equation}
P(0,0)-P(p,0)=\frac{p}{(\gamma-\alpha)+(\delta-\beta)}\left[  \left|
a\right|  ^{2}-\frac{(\gamma-\alpha)}{(\gamma-\alpha)+(\delta-\beta)}\right]
\geq0 \label{Difference1Symmetric}%
\end{equation}
and the difference $\left\{  P(0,0)-P(p,0)\right\}  >0$ when $1\geq\left|
a\right|  ^{2}>\frac{(\gamma-\alpha)}{(\gamma-\alpha)+(\delta-\beta)}$. In
this range of $\left|  a\right|  ^{2}$ the equilibrium $p^{\star}=0$ is a pure
ESS. However, when $\left|  a\right|  ^{2}=\frac{(\gamma-\alpha)}%
{(\gamma-\alpha)+(\delta-\beta)}$ we have the difference $\left\{
P(0,0)-P(p,0)\right\}  $ identically zero. The strategy $p^{\star}=0$ can be
an ESS if%

\begin{align}
&  P(0,p)-P(p,p)\nonumber\\
&  =p\left\{  (\gamma-\alpha)+(\delta-\beta)\right\}  \left\{  \left|
a\right|  ^{2}-\frac{(1-p)(\gamma-\alpha)+p(\delta-\beta)}{(\gamma
-\alpha)+(\delta-\beta)}\right\}  >0
\end{align}
that can be written as%

\begin{equation}
P(0,p)-P(p,p)=p\left\{  (\gamma-\alpha)+(\delta-\beta)\right\}  \left\{
\left|  a\right|  ^{2}-\digamma\right\}  >0
\end{equation}
where $\frac{(\gamma-\alpha)}{(\gamma-\alpha)+(\delta-\beta)}\leq\digamma
\leq\frac{(\delta-\beta)}{(\gamma-\alpha)+(\delta-\beta)}$ when $0\leq
p\leq1.$ The strategy $p^{\star}=0$ can be an ESS only when $\left|  a\right|
^{2}>\frac{(\delta-\beta)}{(\gamma-\alpha)+(\delta-\beta)}$ which is not
possible because $\left|  a\right|  ^{2}$ is fixed at $\frac{(\gamma-\alpha
)}{(\gamma-\alpha)+(\delta-\beta)}.$ Therefore the strategy $p^{\star}=0$ is
an ESS for $1\geq\left|  a\right|  ^{2}>\frac{(\gamma-\alpha)}{(\gamma
-\alpha)+(\delta-\beta)}$ and for $\left|  a\right|  ^{2}=\frac{(\gamma
-\alpha)}{(\gamma-\alpha)+(\delta-\beta)}$ this NE becomes unstable. The
classical game is obtained by taking $\left|  a\right|  ^{2}=1$ for which
$p^{\star}=0$ is an ESS or a stable NE. However this NE does not remain stable
for $\left|  a\right|  ^{2}=\frac{(\gamma-\alpha)}{(\gamma-\alpha
)+(\delta-\beta)}$ which corresponds to an entangled initial state; though the
NE remains intact in both forms of the game.

\paragraph{Case $p^{\star}=1:$}

Similar to the last case the NE condition for the strategy $p^{\star}=1$ can
be written as%

\begin{equation}
P(1,1)-P(p,1)=\frac{(1-p)}{(\gamma-\alpha)+(\delta-\beta)}\left[  -\left|
a\right|  ^{2}+\frac{(\delta-\beta)}{(\gamma-\alpha)+(\delta-\beta)}\right]
\geq0\text{.} \label{Difference2Symmetric}%
\end{equation}
Now $p^{\star}=1$ is a pure ESS for $0\leq\left|  a\right|  ^{2}<\frac
{(\delta-\beta)}{(\gamma-\alpha)+(\delta-\beta)}$. For $\left|  a\right|
^{2}=\frac{(\delta-\beta)}{(\gamma-\alpha)+(\delta-\beta)}$ the difference
$\left\{  P(1,1)-P(p,1)\right\}  $ becomes identically zero. The strategy
$p^{\star}=1$ is an ESS when%

\begin{align}
&  P(1,p)-P(p,p)\nonumber\\
&  =(1-p)\left\{  (\gamma-\alpha)+(\delta-\beta)\right\}  \left\{  -\left|
a\right|  ^{2}+\frac{(1-p)(\gamma-\alpha)+p(\delta-\beta)}{(\gamma
-\alpha)+(\delta-\beta)}\right\}  >0\text{.}\nonumber\\
&
\end{align}
It is possible only if $\left|  a\right|  ^{2}<\frac{(\gamma-\alpha)}%
{(\gamma-\alpha)+(\delta-\beta)}.$ Therefore the strategy $p^{\star}=1$ is a
stable NE (ESS) for $0\leq\left|  a\right|  ^{2}<\frac{(\delta-\beta)}%
{(\gamma-\alpha)+(\delta-\beta)}.$ It is not stable classically (i.e. for
$\left|  a\right|  ^{2}=1$) but becomes stable for an entangled initial state.

\paragraph{Case $p^{\star}=\frac{(\delta-\beta)\left|  a\right|  ^{2}%
-(\gamma-\alpha)\left|  b\right|  ^{2}}{(\delta-\beta)-(\gamma-\alpha)}:$}

In case of the mixed strategy:%

\begin{equation}
p^{\star}=\frac{(\delta-\beta)\left|  a\right|  ^{2}-(\gamma-\alpha)\left|
b\right|  ^{2}}{(\delta-\beta)-(\gamma-\alpha)} \label{MixedStrategySmmetric}%
\end{equation}
the NE condition (\ref{General2PlayerDifference1}) turns into $P(p^{\star
},p^{\star})-P(p,p^{\star})=0$. The mixed strategy
(\ref{MixedStrategySmmetric}) can be an ESS if%

\begin{align}
&  P(p^{\star},p)-P(p,p)\nonumber\\
&  =(p^{\star}-p)\left[  -\left|  a\right|  ^{2}(\delta-\beta)+\left|
b\right|  ^{2}(\gamma-\alpha)+p\left\{  (\delta-\beta)-(\gamma-\alpha
)\right\}  \right]  >0\nonumber\\
&  \label{Difference3Symmetric}%
\end{align}
for all $p\neq p^{\star}$. Write now the strategy $p$ as $p=p^{\star
}+\bigtriangleup$. For the mixed strategy (\ref{MixedStrategySmmetric}) the
payoff difference of the Eq. (\ref{Difference3Symmetric}) is reduced to%

\begin{equation}
P(p^{\star},p)-P(p,p)=-\bigtriangleup^{2}\left\{  (\delta-\beta)-(\gamma
-\alpha)\right\}  \text{.}%
\end{equation}
Hence, for the game defined in the conditions (\ref{GameDefinition}), the
mixed strategy $p^{\star}=\frac{(\delta-\beta)\left|  a\right|  ^{2}%
-(\gamma-\alpha)\left|  b\right|  ^{2}}{(\delta-\beta)-(\gamma-\alpha)}$
cannot be an ESS, though it can be a NE of the symmetric game.

It is to be pointed out that above considerations apply when the game is
played with the initial state (\ref{IniStatGen2Player}).

To find examples of symmetric quantum games, where evolutionary stability of
the mixed strategies may also be affected by controlling the initial states,
the number of players are now increased from two to three.

\subsubsection{$2\times2\times2$ symmetric games}

In extending the two-player scheme to a three-player case, we assume that
three players $A,B,$ and $C$ play their strategies by applying the identity
operator $\hat{I}$ with the probabilities $p,q$ and $r$ respectively on the
initial state $\left|  \psi_{in}\right\rangle $. Therefore, they apply the
operator $\hat{\sigma}_{x}$ with the probabilities $(1-p),(1-q)$ and $(1-r)$
respectively. The final state becomes%

\begin{equation}
\rho_{fin}=\underset{\hat{U}=\hat{I},\hat{\sigma}_{x}}{\sum}\Pr(\hat{U}%
_{A})\Pr(\hat{U}_{B})\Pr(\hat{U}_{C})\left[  \hat{U}_{A}\otimes\hat{U}%
_{B}\otimes\hat{U}_{C}\rho_{in}\hat{U}_{A}^{\dagger}\otimes\hat{U}%
_{B}^{\dagger}\otimes\hat{U}_{C}^{\dagger}\right]
\end{equation}
where the $8$ basis vectors are $\left|  S_{i}S_{j}S_{k}\right\rangle $, for
$i,j,k=1,2$. Again we use initial quantum state in the form $\left|  \psi
_{in}\right\rangle =a\left|  S_{1}S_{1}S_{1}\right\rangle +b\left|  S_{2}%
S_{2}S_{2}\right\rangle $, where $\left|  a\right|  ^{2}+\left|  b\right|
^{2}=1$. It is a quantum state in $2\otimes2\otimes2$ dimensional Hilbert
space
\index{Hilbert space}
that can be prepared from a system of three two-state quantum systems or
qubits. Similar to the two-player case, the payoff operators
\index{Payoff operators}
for the players $A,$ $B,$ and $C$ can be defined as%

\begin{align}
&  (P_{A,B,C})_{oper}=\nonumber\\
&  \alpha_{1},\beta_{1},\eta_{1}\left|  S_{1}S_{1}S_{1}\right\rangle
\left\langle S_{1}S_{1}S_{1}\right|  +\alpha_{2},\beta_{2},\eta_{2}\left|
S_{2}S_{1}S_{1}\right\rangle \left\langle S_{2}S_{1}S_{1}\right|  +\nonumber\\
&  \alpha_{3},\beta_{3},\eta_{3}\left|  S_{1}S_{2}S_{1}\right\rangle
\left\langle S_{1}S_{2}S_{1}\right|  +\alpha_{4},\beta_{4},\eta_{4}\left|
S_{1}S_{1}S_{2}\right\rangle \left\langle S_{1}S_{1}S_{2}\right|  +\nonumber\\
&  \alpha_{5},\beta_{5},\eta_{5}\left|  S_{1}S_{2}S_{2}\right\rangle
\left\langle S_{1}S_{2}S_{2}\right|  +\alpha_{6},\beta_{6},\eta_{6}\left|
S_{2}S_{1}S_{2}\right\rangle \left\langle S_{2}S_{1}S_{2}\right|  +\nonumber\\
&  \alpha_{7},\beta_{7},\eta_{7}\left|  S_{2}S_{2}S_{1}\right\rangle
\left\langle S_{2}S_{2}S_{1}\right|  +\alpha_{8},\beta_{8},\eta_{8}\left|
S_{2}S_{2}S_{2}\right\rangle \left\langle S_{2}S_{2}S_{2}\right|
\end{align}
where $\alpha_{l},\beta_{l},\eta_{l}$ for $1\leq l\leq8$ are $24$ constants of
the matrix of this three-player game. Payoffs to the players $A,B,$ and $C$
are then obtained as mean values of these operators i.e. $P_{A,B,C}%
(p,q,r)=$Tr$\left[  (P_{A,B,C})_{oper}\rho_{fin}\right]  $.

Here, similar to the two-player case, the classical payoffs can be obtained
when $\left|  b\right|  ^{2}=0$. To get a symmetric game we define
$P_{A}(x,y,z)$ as the payoff to player $A$ when players $A$, $B$, and $C$ play
the strategies $x$, $y$, and $z$ respectively. With following relations the
players' payoffs become identity-independent.%

\begin{gather}
P_{A}(x,y,z)=P_{A}(x,z,y)=P_{B}(y,x,z)\nonumber\\
=P_{B}(z,x,y)=P_{C}(y,z,x)=P_{C}(z,y,x)\text{.}
\label{3PlayerSymmetricConditions}%
\end{gather}
The players in the game then become anonymous and their payoffs depend only on
their strategies. The relations (\ref{3PlayerSymmetricConditions}) hold with
the following replacements for $\beta_{i}$ and $\eta_{i}$:%

\begin{align}
\beta_{1}  &  \rightarrow\alpha_{1}\qquad\beta_{2}\rightarrow\alpha_{3}%
\qquad\beta_{3}\rightarrow\alpha_{2}\qquad\beta_{4}\rightarrow\alpha
_{3}\nonumber\\
\beta_{5}  &  \rightarrow\alpha_{6}\qquad\beta_{6}\rightarrow\alpha_{5}%
\qquad\beta_{7}\rightarrow\alpha_{6}\qquad\beta_{8}\rightarrow\alpha
_{8}\nonumber\\
\eta_{1}  &  \rightarrow\alpha_{1}\qquad\eta_{2}\rightarrow\alpha_{3}%
\qquad\eta_{3}\rightarrow\alpha_{3}\qquad\eta_{4}\rightarrow\alpha
_{2}\nonumber\\
\eta_{5}  &  \rightarrow\alpha_{6}\qquad\eta_{6}\rightarrow\alpha_{6}%
\qquad\eta_{7}\rightarrow\alpha_{5}\qquad\eta_{8}\rightarrow\alpha_{8}\text{.}%
\end{align}
Also, it is now necessary that we should have $\alpha_{6}=\alpha_{7}$,
$\alpha_{3}=\alpha_{4}$.

A symmetric game between three players, therefore, can be defined by only six
constants of the payoff matrix
\index{Payoff matrix}%
. These constants can be taken as $\alpha_{1},\alpha_{2},\alpha_{3},\alpha
_{5},\alpha_{6},$ and $\alpha_{8}$. Payoff to a $p$-player, when other two
players play $q$ and $r$, can now be written as $P(p,q,r)$. A symmetric NE
$\overset{\star}{p}$ is now found from the Nash condition $P(\overset{\star
}{p},\overset{\star}{p},\overset{\star}{p})-P(p,\overset{\star}{p}%
,\overset{\star}{p})\geq0$ i.e.%

\begin{gather}
P(\overset{\star}{p},\overset{\star}{p},\overset{\star}{p})-P(p,\overset
{\star}{p},\overset{\star}{p})=(\overset{\star}{p}-p)\text{{\LARGE [}}%
\overset{\star}{p}^{2}(1-2\left|  b\right|  ^{2})(\sigma+\omega-2\eta
)+\nonumber\\
2\overset{\star}{p}\left\{  \left|  b\right|  ^{2}(\sigma+\omega-2\eta
)-\omega+\eta\right\}  +\left\{  \omega-\left|  b\right|  ^{2}(\sigma
+\omega)\right\}  \text{{\LARGE ]}}\geq0
\end{gather}
where $(\alpha_{1}-\alpha_{2})=\sigma,(\alpha_{3}-\alpha_{6})=\eta,$and
$(\alpha_{5}-\alpha_{8})=\omega$.

Three possible NE are found as%

\begin{equation}
\left.
\begin{array}
[c]{c}%
\overset{\star}{p}_{1}=\frac{\left\{  (\omega-\eta)-\left|  b\right|
^{2}(\sigma+\omega-2\eta)\right\}  \pm\sqrt{\left\{  (\sigma+\omega
)^{2}-(2\eta)^{2}\right\}  \left|  b\right|  ^{2}(1-\left|  b\right|
^{2})+(\eta^{2}-\sigma\omega)}}{(1-2\left|  b\right|  ^{2})(\sigma
+\omega-2\eta)}\\%
\begin{array}
[c]{c}%
\overset{\star}{p}_{2}=0\\
\overset{\star}{p}_{3}=1
\end{array}
\end{array}
\right\}  \text{.}%
\end{equation}
It is observed that the mixed NE $\overset{\star}{p_{1}}$ makes the difference
$\left\{  P(\overset{\star}{p},\overset{\star}{p},\overset{\star}%
{p})-P(p,\overset{\star}{p},\overset{\star}{p})\right\}  $ identically zero
and two values for $\overset{\star}{p}_{1}$ can be found for a given $\left|
b\right|  ^{2}$. Apart from $\overset{\star}{p}_{1}$ the other two NE (i.e.
$\overset{\star}{p}_{2}$ and $\overset{\star}{p}_{3}$) are pure strategies.
Also now $\overset{\star}{p}_{1}$ comes out a NE without imposing further
restrictions on the matrix of the symmetric three-player game. However, the
pure strategies $\overset{\star}{p}_{2}$ and $\overset{\star}{p}_{3}$ can be
NE when further restriction are imposed on the matrix of the game. For
example, $\overset{\star}{p}_{3}$ can be a NE provided $\sigma\geq
(\omega+\sigma)\left|  b\right|  ^{2}$ for all $\left|  b\right|  ^{2}%
\in\lbrack0,1]$. Similarly $\overset{\star}{p}_{2}$ can be NE when $\omega
\leq(\omega+\sigma)\left|  b\right|  ^{2}$.

Now we address the question: How evolutionary stability of these three NE can
be affected while playing the game via initial quantum states given in the
following form?%

\begin{equation}
\left|  \psi_{in}\right\rangle =a\left|  S_{1}S_{1}S_{1}\right\rangle
+b\left|  S_{2}S_{2}S_{2}\right\rangle \text{.} \label{IniStatSymm3Player}%
\end{equation}
For the two-player asymmetric game of BoS we showed that out of three NE only
two can be evolutionarily stable. In classical evolutionary game theory the
concept of an ESS is well-known \cite{MarkBroom1,BroomMutiplayer} to be
extendable to multi-player case. When mutants are allowed to play only one
strategy the definition of an ESS for the three-player case is written as
\cite{MarkBroom1}%

\begin{align}
1.\text{ \ \ \ }P(p,p,p)  &  >P(q,p,p)\nonumber\\
2.\text{ If }P(p,p,p)  &  =P(q,p,p)\text{ then }P(p,q,p)>P(q,q,p)\text{.}%
\end{align}
Here $p$ is a NE if it satisfies the condition $1$ against all $q\neq p$. For
our case the ESS conditions for the pure strategies $\overset{\star}{p}_{2}$
and $\overset{\star}{p}_{3}$ can be written as follows. For example
$\overset{\star}{p}_{2}=0$ is an ESS when%

\begin{align}
1.\text{ \ \ \ }\sigma\left|  b\right|  ^{2}  &  >\omega\left|  a\right|
^{2}\nonumber\\
2.\text{ If }\sigma\left|  b\right|  ^{2}  &  =\omega\left|  a\right|
^{2}\text{ then }-\eta q^{2}(\left|  a\right|  ^{2}-\left|  b\right|  ^{2})>0
\label{Cond3PlayerSymme1}%
\end{align}
where $1$ is NE condition for the strategy $\overset{\star}{p}_{2}=0$.
Similarly, $\overset{\star}{p}_{3}=1$ is an ESS when%

\begin{align}
1.\text{ \ \ \ }\sigma\left|  a\right|  ^{2}  &  >\omega\left|  b\right|
^{2}\nonumber\\
2.\text{ If }\sigma\left|  a\right|  ^{2}  &  =\omega\left|  b\right|
^{2}\text{ then }\eta(1-q)^{2}(\left|  a\right|  ^{2}-\left|  b\right|
^{2})>0 \label{Conditions3PlayerSymm}%
\end{align}
and both the pure strategies $\overset{\star}{p}_{2}$ and $\overset{\star}%
{p}_{3}$ are ESSs when $\left|  a\right|  ^{2}=\left|  b\right|  ^{2}$. The
conditions (\ref{Conditions3PlayerSymm}) can also be written as
\begin{align}
1.\text{ \ \ \ }\sigma &  >(\omega+\sigma)\left|  b\right|  ^{2}\nonumber\\
2.\text{ If }\sigma &  =\left|  b\right|  ^{2}(\omega+\sigma)\text{ then
}\frac{\gamma(\omega-\sigma)}{(\omega+\sigma)}>0\text{.}%
\end{align}
For the strategy $\overset{\star}{p}_{2}=0$ the ESS conditions
(\ref{Cond3PlayerSymme1}) reduce to%

\begin{align}
1.\text{ \ \ \ \ }\omega &  <(\omega+\sigma)\left|  b\right|  ^{2}\nonumber\\
2.\text{ If \ }\omega &  =\left|  b\right|  ^{2}(\omega+\sigma)\text{ then
}\frac{\gamma(\omega-\sigma)}{(\omega+\sigma)}>0
\end{align}
Examples of three-player symmetric games are easy to find for which a pure
strategy is a NE for the whole range $\left|  b\right|  ^{2}\in\lbrack0,1]$,
but it is not an ESS for some particular value of $\left|  b\right|  ^{2}$. An
example of a class of such games is for which $\sigma=0$, $\omega<0$, and
$\eta\leq0$. In this class the strategy $\overset{\star}{p}_{2}=0$ is a NE for
all $\left|  b\right|  ^{2}\in\lbrack0,1]$ but not an ESS when $\left|
b\right|  ^{2}=1$.

Apart from the pure strategies, the mixed strategy equilibrium $\overset
{\star}{p}_{1}$ forms the most interesting case. It makes the payoff
difference $\left\{  P(\overset{\star}{p_{1}},\overset{\star}{p_{1}}%
,\overset{\star}{p_{1}})-P(p,\overset{\star}{p_{1}},\overset{\star}{p_{1}%
})\right\}  $ identically zero for every strategy $p$. The strategy
$\overset{\star}{p_{1}} $ is an ESS when $\left\{  P(\overset{\star}{p_{1}%
},q,\overset{\star}{p_{1}})-P(q,q,\overset{\star}{p_{1}})\right\}  >0$ but%

\begin{align}
&  P(\overset{\star}{p_{1}},q,\overset{\star}{p_{1}})-P(q,q,\overset{\star
}{p_{1}})\nonumber\\
&  =\pm(\overset{\star}{p_{1}}-q)^{2}\sqrt{\left\{  (\sigma+\omega)^{2}%
-(2\eta)^{2}\right\}  \left|  b\right|  ^{2}(1-\left|  b\right|  ^{2}%
)+(\eta^{2}-\sigma\omega)}\text{.} \label{3PlayerSqr}%
\end{align}
Therefore, out of the two possible roots $(\overset{\star}{p_{1}})_{1}$ and
$(\overset{\star}{p_{1}})_{2}$ of the quadratic equation\footnote{These roots
make the difference $\left\{  P(\overset{\star}{p_{1}},q,\overset{\star}%
{p_{1}})-P(q,q,\overset{\star}{p_{1}})\right\}  $ greater than and less than
zero, respectively.}:%

\begin{gather}
\overset{\star}{p_{1}}^{2}(1-2\left|  b\right|  ^{2})(\sigma+\omega
-2\eta)+\nonumber\\
2\overset{\star}{p_{1}}\left\{  \left|  b\right|  ^{2}(\sigma+\omega
-2\eta)-\omega+\eta\right\}  +\left\{  \omega-\left|  b\right|  ^{2}%
(\sigma+\omega)\right\}  =0 \label{DefQuadEqForp1Star}%
\end{gather}
only $(\overset{\star}{p_{1}})_{1}$ can exist as an ESS. When the square root
term in the equation (\ref{3PlayerSqr}) becomes zero we have only one mixed
NE, which is not an ESS. Hence, out of four possible NE in this three-player
game only three can be ESSs.

An interesting class of three-player games is the one for which $\eta
^{2}=\sigma\omega$. For these games the mixed NE are%

\begin{equation}
\overset{\star}{p_{1}}=\frac{\left\{  (w-\eta)-\left|  b\right|  ^{2}%
(\sigma+\omega-2\eta)\right\}  \pm\left|  a\right|  \left|  b\right|  \left|
\sigma-\omega\right|  }{(1-2\left|  b\right|  ^{2})(\sigma+\omega-2\eta)}%
\end{equation}
and, when played classically, we can get only one mixed NE that is not an ESS.
However for all $\left|  b\right|  ^{2}$, different from zero, we generally
obtain two NE out of which one can be an ESS.

Similar to the two-player case, the equilibria in a three-player symmetric
game where quantization affects evolutionary stability, are the ones that
survive for two initial states, one of which is a product state and
corresponds to the classical game. Suppose $\overset{\star}{p_{1}}$ remains a
NE for $\left|  b\right|  ^{2}=0$ and some other non-zero $\left|  b\right|
^{2}$. It is possible when $(\sigma-\omega)(2\overset{\star}{p_{1}}-1)=0$.
Alternatively, the strategy $\overset{\star}{p}=\frac{1}{2}$ remains a NE for
all $\left|  b\right|  ^{2}\in\lbrack0,1]$. It reduces the defining quadratic
equation (\ref{DefQuadEqForp1Star}) for $\overset{\star}{p_{1}}$ to
$\sigma+\omega+2\eta=0$ and makes the difference $\left\{  P(\overset{\star
}{p_{1}},q,\overset{\star}{p_{1}})-P(q,q,\overset{\star}{p_{1}})\right\}  $
independent of $\left|  b\right|  ^{2}$. Therefore the strategy $\overset
{\star}{p}=\frac{1}{2}$, even when it is retained as an equilibrium for all
$\left|  b\right|  ^{2}\in\lbrack0,1]$, cannot be an ESS in only one version
of the symmetric three-player game. For the second possibility $\sigma=\omega$
the defining equation for $\overset{\star}{p_{1}}$ is reduced to%

\begin{equation}
(1-2\left|  b\right|  ^{2})\left\{  \overset{\star}{p_{1}}-\frac{(\eta
-\sigma)-\sqrt{\eta^{2}-\sigma^{2}}}{2(\eta-\sigma)}\right\}  \left\{
\overset{\star}{p_{1}}-\frac{(\eta-\sigma)+\sqrt{\eta^{2}-\sigma^{2}}}%
{2(\eta-\sigma)}\right\}  =0
\end{equation}
for which%

\begin{equation}
P(\overset{\star}{p_{1}},q,\overset{\star}{p_{1}})-P(q,q,\overset{\star}%
{p_{1}})=\pm2(\overset{\star}{p_{1}}-q)^{2}\left|  \left|  b\right|
^{2}-\frac{1}{2}\right|  \sqrt{\eta^{2}-\sigma^{2}}\text{.}%
\end{equation}
Here the difference $\left\{  P(\overset{\star}{p_{1}},q,\overset{\star}%
{p_{1}})-P(q,q,\overset{\star}{p_{1}})\right\}  $ still depends on $\left|
b\right|  ^{2}$ and becomes zero for $\left|  b\right|  ^{2}=1/2$.

Hence, for the class of games for which $\sigma=\omega$ and $\eta>\sigma$, one
of the mixed strategies $(\overset{\star}{p_{1}})_{1},(\overset{\star}{p_{1}%
})_{2}$ remains a NE for all $\left|  b\right|  ^{2}\in\lbrack0,1]$ but not an
ESS when $\left|  b\right|  ^{2}=1/2$. In this class of three-player quantum
games the evolutionary stability of a mixed strategy can, therefore, be
changed while the game is played using initial quantum states in the form
(\ref{IniStatSymm3Player}).

\subsubsection{Rock-Scissors-Paper game}

Rock-Scissors-Paper (RSP)
\index{Rock-Scissors-Paper (RSP) game}
is a game for two players that is typically played using the players' hands.
This game has been played for long as a children's pastime or as an
odd-man-out selection process. The players opposite each others, tap their
fist in their open palm three times (saying Rock, Scissors, Paper) and then
show one of three possible gestures. The Rock wins against the scissors
(crushes it) but looses against the paper (is wrapped into it). The Scissors
wins against the paper (cuts it) but looses against the rock (is crushed by
it). The Paper wins against the rock (wraps it) but looses against the
scissors (is cut by it). The game is also played in nature like many other
games. Lizards in the Coast Range of California play this game
\cite{LizardGame} using three alternative male strategies locked in an
ecological never ending process from which there seems little escape.

In a slightly modified version of the RSP game both players get a small
premium $\epsilon$\ for a draw. This game can be represented by the payoff
matrix
\index{Payoff matrix}%
:%

\begin{equation}%
\begin{array}
[c]{c}%
R\\
S\\
P
\end{array}
\overset{%
\begin{array}
[c]{ccc}%
R & S & P
\end{array}
}{\left(
\begin{array}
[c]{ccc}%
-\epsilon & 1 & -1\\
-1 & -\epsilon & 1\\
1 & -1 & -\epsilon
\end{array}
\right)  } \label{RSPMatrix}%
\end{equation}
where $-1<\epsilon\leq0$. The matrix of the usual game is obtained when
$\epsilon$ is zero.

One cannot win if one's opponent knew which strategy was going to be picked.
For example, picking Rock consistently all the opponent needs to do is pick
Paper and s/he would win. Players find soon that in case predicting opponent's
strategy is not possible the best strategy is to pick Rock, Scissors, or Paper
at random. In other words, the player selects Rock, Scissors, or Paper with a
probability of $1/3$. In case opponent's strategy is predictable picking a
strategy at random with a probability of $1/3$ is \emph{not} the best thing to
do unless the opponent does the same \cite{Prestwich}.

Analysis \cite{Weibull} of the modified RSP game of matrix (\ref{RSPMatrix})
shows that its NE consists of playing each of the three different pure
strategies with a fixed equilibrial probability $1/3$. However it is not an
ESS because $\epsilon$ is negative.

Here we want to study the effects of quantization on evolutionary stability
for the modified RSP game. The game is different, from others considered
earlier, because classically each player now possesses three pure strategies
instead of two. A classical mixed NE exists which is not an ESS. Our
motivation is to explore the possibility that the classical mixed NE becomes
an ESS for some quantum form of the game.

\paragraph{Quantization of Rock-Scissors-Paper game:}

Using simpler notation: $R\sim1,$ $S\sim2,$ $P\sim3$ we quantize this game via
MW scheme
\index{MW quantization scheme}
\cite{Marinatto1}. We assume the two players are in possession of three
unitary and Hermitian operators
\index{Hermitian operator}
$\hat{I},$ $\hat{C}$ and $\hat{D}$ defined as follows.%

\begin{align}
\hat{I}\left|  1\right\rangle  &  =\left|  1\right\rangle \text{,}\qquad
\hat{C}\left|  1\right\rangle =\left|  3\right\rangle \text{,}\qquad\hat
{D}\left|  1\right\rangle =\left|  2\right\rangle \nonumber\\
\hat{I}\left|  2\right\rangle  &  =\left|  2\right\rangle \text{,}\qquad
\hat{C}\left|  2\right\rangle =\left|  2\right\rangle \text{,}\qquad\hat
{D}\left|  2\right\rangle =\left|  1\right\rangle \nonumber\\
\hat{I}\left|  3\right\rangle  &  =\left|  3\right\rangle \text{,}\qquad
\hat{C}\left|  3\right\rangle =\left|  1\right\rangle \text{,}\qquad\hat
{D}\left|  3\right\rangle =\left|  3\right\rangle
\end{align}
where $\hat{C}^{\dagger}=\hat{C}=\hat{C}^{-1}$ and $\hat{D}^{\dagger}=\hat
{D}=\hat{D}^{-1}$ and $\hat{I}$ is the identity operator.

Consider a general two-player payoff matrix
\index{Payoff matrix}
when each player has three strategies:%

\begin{equation}%
\begin{array}
[c]{c}%
1\\
2\\
3
\end{array}
\overset{%
\begin{array}
[c]{ccccccc}%
1 &  &  & 2 &  &  & 3
\end{array}
}{\left(
\begin{array}
[c]{ccc}%
(\alpha_{11},\beta_{11}) & (\alpha_{12},\beta_{12}) & (\alpha_{13},\beta
_{13})\\
(\alpha_{21},\beta_{21}) & (\alpha_{22},\beta_{22}) & (\alpha_{23},\beta
_{23})\\
(\alpha_{31},\beta_{31}) & (\alpha_{32},\beta_{32}) & (\alpha_{33},\beta_{33})
\end{array}
\right)  } \label{Gen3StrategyMatrix}%
\end{equation}
where $\alpha_{ij}$ and $\beta_{ij}$ are the payoffs to Alice and Bob,
respectively, when Alice plays $i$ and Bob plays $j$ and $1\leq i,j\leq3$.
Suppose Alice and Bob apply the operators $\hat{C}$, $\hat{D}$, and $\hat{I}$
with the probabilities $p$, $p_{1}$, $(1-p-p_{1})$ and $q$, $q_{1}$,
$(1-q-q_{1})$ respectively. The initial state of the game is $\rho_{in}$.
Alice's move changes the state changes to%

\begin{equation}
\overset{A}{\rho_{in}}=(1-p-p_{1})\hat{I}_{A}\rho_{in}\hat{I}_{A}^{\dagger
}+p\hat{C}_{A}\rho_{in}\hat{C}_{A}^{\dagger}+p_{1}\hat{D}_{A}\rho_{in}\hat
{D}_{A}^{\dagger}\text{.}%
\end{equation}
The final state, after Bob too has played his move, is%

\begin{equation}
\overset{A,B}{\rho_{f}}=(1-q-q_{1})\hat{I}_{B}\overset{A}{\rho_{in}}\hat
{I}_{B}^{\dagger}+q\hat{C}_{B}\overset{A}{\rho_{in}}\hat{C}_{B}^{\dagger
}+q_{1}\hat{D}_{B}\overset{A}{\rho_{in}}\hat{D}_{B}^{\dagger}\text{.}%
\end{equation}
This state can be written as%

\begin{align}
\overset{A,B}{\rho_{f}}  &  =(1-p-p_{1})(1-q-q_{1})\left\{  \hat{I}_{A}%
\otimes\hat{I}_{B}\rho_{in}\hat{I}_{A}^{\dagger}\otimes\hat{I}_{B}^{\dagger
}\right\}  +p(1-q-q_{1})\times\nonumber\\
&  \left\{  \hat{C}_{A}\otimes\hat{I}_{B}\rho_{in}\hat{C}_{A}^{\dagger}%
\otimes\hat{I}_{B}^{\dagger}\right\}  +p_{1}(1-q-q_{1})\left\{  \hat{D}%
_{A}\otimes\hat{I}_{B}\rho_{in}\hat{D}_{A}^{\dagger}\otimes\hat{I}%
_{B}^{\dagger}\right\}  +\nonumber\\
&  (1-p-p_{1})q\left\{  \hat{I}_{A}\otimes\hat{C}_{B}\rho_{in}\hat{I}%
_{A}^{\dagger}\otimes\hat{C}_{B}^{\dagger}\right\}  +pq\left\{  \hat{C}%
_{A}\otimes\hat{C}_{B}\rho_{in}\hat{C}_{A}^{\dagger}\otimes\hat{C}%
_{B}^{\dagger}\right\}  +\nonumber\\
&  p_{1}q\left\{  \hat{D}_{A}\otimes\hat{C}_{B}\rho_{in}\hat{D}_{A}^{\dagger
}\otimes\hat{C}_{B}^{\dagger}\right\}  +(1-p-p_{1})q_{1}\left\{  \hat{I}%
_{A}\otimes\hat{D}_{B}\rho_{in}\hat{I}_{A}^{\dagger}\otimes\hat{D}%
_{B}^{\dagger}\right\} \nonumber\\
&  +pq_{1}\left\{  \hat{C}_{A}\otimes\hat{D}_{B}\rho_{in}\hat{C}_{A}^{\dagger
}\otimes\hat{D}_{B}^{\dagger}\right\}  +p_{1}q_{1}\left\{  \hat{D}_{A}%
\otimes\hat{D}_{B}\rho_{in}\hat{D}_{A}^{\dagger}\otimes\hat{D}_{B}^{\dagger
}\right\}  \text{.}%
\end{align}
The nine basis vectors of initial quantum state with three pure classical
strategies are $\left|  ij\right\rangle $ for $i,j=1,2,3$. We consider the
initial state to be a pure quantum state of two qutrits i.e.%

\begin{equation}
\left|  \psi_{in}\right\rangle =\underset{i,j=1,2,3}{\sum}c_{ij}\left|
ij\right\rangle \text{, \ \ \ \ where }\underset{i,j=1,2,3}{\sum}\left|
c_{ij}\right|  ^{2}=1\text{.} \label{RSPIniStat}%
\end{equation}
The payoff operators
\index{Payoff operators}
for Alice and Bob are \cite{Marinatto1}%

\begin{align}
(P_{A,B})_{oper}  &  =(\alpha,\beta)_{11}\left|  11\right\rangle \left\langle
11\right|  +(\alpha,\beta)_{12}\left|  12\right\rangle \left\langle 12\right|
+(\alpha,\beta)_{13}\left|  13\right\rangle \left\langle 13\right|
+\nonumber\\
&  (\alpha,\beta)_{21}\left|  21\right\rangle \left\langle 21\right|
+(\alpha,\beta)_{22}\left|  22\right\rangle \left\langle 22\right|
+(\alpha,\beta)_{23}\left|  23\right\rangle \left\langle 23\right|
+\nonumber\\
&  (\alpha,\beta)_{31}\left|  31\right\rangle \left\langle 31\right|
+(\alpha,\beta)_{32}\left|  32\right\rangle \left\langle 32\right|
+(\alpha,\beta)_{33}\left|  33\right\rangle \left\langle 33\right|
\text{.}\nonumber\\
&
\end{align}
The players' payoffs are then%

\begin{equation}
P_{A,B}=\text{Tr}[\left\{  (P_{A,B})_{oper}\right\}  \overset{A,B}{\rho_{f}}]
\label{3StrategyPayoff}%
\end{equation}
Payoff to Alice, for example, can be written as%

\begin{equation}
P_{A}=\Phi\Omega\Upsilon^{T} \label{RSPPayoffAlice}%
\end{equation}
where $T$ is for transpose, and the matrices $\Phi,$ $\Omega,$ and $\Upsilon$ are%

\begin{align}
\Phi &  =[%
\begin{array}
[c]{ccc}%
(1-p-p_{1})(1-q-q_{1}) & p(1-q-q_{1}) & p_{1}(1-q-q_{1})
\end{array}
\nonumber\\
&
\begin{array}
[c]{cccccc}%
(1-p-p_{1})q & pq & p_{1}q & (1-p-p_{1})q_{1} & pq_{1} & p_{1}q_{1}%
\end{array}
]\nonumber\\
\Upsilon &  =[%
\begin{array}
[c]{ccccccccc}%
\alpha_{11} & \alpha_{12} & \alpha_{13} & \alpha_{21} & \alpha_{22} &
\alpha_{23} & \alpha_{31} & \alpha_{32} & \alpha_{33}%
\end{array}
]\nonumber\\
\Omega &  =\left[
\begin{array}
[c]{ccccccccc}%
\left|  c_{11}\right|  ^{2} & \left|  c_{12}\right|  ^{2} & \left|
c_{13}\right|  ^{2} & \left|  c_{21}\right|  ^{2} & \left|  c_{22}\right|
^{2} & \left|  c_{23}\right|  ^{2} & \left|  c_{31}\right|  ^{2} & \left|
c_{32}\right|  ^{2} & \left|  c_{33}\right|  ^{2}\\
\left|  c_{31}\right|  ^{2} & \left|  c_{32}\right|  ^{2} & \left|
c_{33}\right|  ^{2} & \left|  c_{21}\right|  ^{2} & \left|  c_{22}\right|
^{2} & \left|  c_{23}\right|  ^{2} & \left|  c_{11}\right|  ^{2} & \left|
c_{12}\right|  ^{2} & \left|  c_{13}\right|  ^{2}\\
\left|  c_{21}\right|  ^{2} & \left|  c_{22}\right|  ^{2} & \left|
c_{23}\right|  ^{2} & \left|  c_{11}\right|  ^{2} & \left|  c_{12}\right|
^{2} & \left|  c_{13}\right|  ^{2} & \left|  c_{31}\right|  ^{2} & \left|
c_{32}\right|  ^{2} & \left|  c_{33}\right|  ^{2}\\
\left|  c_{13}\right|  ^{2} & \left|  c_{12}\right|  ^{2} & \left|
c_{11}\right|  ^{2} & \left|  c_{23}\right|  ^{2} & \left|  c_{22}\right|
^{2} & \left|  c_{21}\right|  ^{2} & \left|  c_{33}\right|  ^{2} & \left|
c_{32}\right|  ^{2} & \left|  c_{31}\right|  ^{2}\\
\left|  c_{33}\right|  ^{2} & \left|  c_{32}\right|  ^{2} & \left|
c_{31}\right|  ^{2} & \left|  c_{23}\right|  ^{2} & \left|  c_{22}\right|
^{2} & \left|  c_{21}\right|  ^{2} & \left|  c_{13}\right|  ^{2} & \left|
c_{12}\right|  ^{2} & \left|  c_{11}\right|  ^{2}\\
\left|  c_{23}\right|  ^{2} & \left|  c_{22}\right|  ^{2} & \left|
c_{21}\right|  ^{2} & \left|  c_{13}\right|  ^{2} & \left|  c_{12}\right|
^{2} & \left|  c_{11}\right|  ^{2} & \left|  c_{33}\right|  ^{2} & \left|
c_{32}\right|  ^{2} & \left|  c_{31}\right|  ^{2}\\
\left|  c_{12}\right|  ^{2} & \left|  c_{11}\right|  ^{2} & \left|
c_{13}\right|  ^{2} & \left|  c_{22}\right|  ^{2} & \left|  c_{21}\right|
^{2} & \left|  c_{23}\right|  ^{2} & \left|  c_{32}\right|  ^{2} & \left|
c_{31}\right|  ^{2} & \left|  c_{33}\right|  ^{2}\\
\left|  c_{32}\right|  ^{2} & \left|  c_{31}\right|  ^{2} & \left|
c_{33}\right|  ^{2} & \left|  c_{22}\right|  ^{2} & \left|  c_{21}\right|
^{2} & \left|  c_{23}\right|  ^{2} & \left|  c_{12}\right|  ^{2} & \left|
c_{11}\right|  ^{2} & \left|  c_{13}\right|  ^{2}\\
\left|  c_{22}\right|  ^{2} & \left|  c_{21}\right|  ^{2} & \left|
c_{23}\right|  ^{2} & \left|  c_{12}\right|  ^{2} & \left|  c_{11}\right|
^{2} & \left|  c_{13}\right|  ^{2} & \left|  c_{32}\right|  ^{2} & \left|
c_{31}\right|  ^{2} & \left|  c_{33}\right|  ^{2}%
\end{array}
\right]  \text{.}\nonumber\\
&  \label{RSPAliceSigma}%
\end{align}
The payoff (\ref{RSPPayoffAlice}) corresponds to the matrix
(\ref{Gen3StrategyMatrix}). Payoffs in classical mixed strategy game can be
obtained from Eq. (\ref{3StrategyPayoff}) for the initial state $\left|
\psi_{in}\right\rangle =\left|  11\right\rangle $. The game is symmetric when
$\alpha_{ij}=\beta_{ji}$ in the matrix (\ref{Gen3StrategyMatrix}). The quantum
game played using the quantum state (\ref{RSPIniStat}) is symmetric when
$\left|  c_{ij}\right|  ^{2}=\left|  c_{ji}\right|  ^{2}$ for all constants
$c_{ij}$ in the state (\ref{RSPIniStat}). These two conditions together
guarantee a symmetric quantum game. The players' payoffs $P_{A}$, $P_{B}$ then
do not need a subscript and we can simply use $P(p,q)$ to denote the payoff to
the $p$-player against the $q$-player.

The question of evolutionary stability in quantized RSP game is addressed below.

\paragraph{Analysis of evolutionary stability:}

Assume a strategy is defined by a pair of numbers $(p,p_{1})$ for players
playing the quantized RSP game. These numbers are the probabilities with which
the player applies the operators $\hat{C}$ and $\hat{D}$. The identity
operator $\hat{I}$ is, then, applied with probability $(1-p-p_{1})$. Similar
to the conditions a) and b) in Eq. (\ref{ESSDef}), the conditions making a
strategy $(p^{\star},p_{1}^{\star})$ an ESS can be written as \cite{Smith
Price,Weibull}
\begin{align}
1\text{. \ \ \ }P\{(p^{\star},p_{1}^{\star}),(p^{\star},p_{1}^{\star})\}  &
>P\{(p,p_{1}),(p^{\star},p_{1}^{\star})\}\nonumber\\
2\text{. if }P\{(p^{\star},p_{1}^{\star}),(p^{\star},p_{1}^{\star})\}  &
=P\{(p,p_{1}),(p^{\star},p_{1}^{\star})\}\text{ then}\nonumber\\
P\{(p^{\star},p_{1}^{\star}),(p,p_{1})\}  &  >P\{(p,p_{1}),(p,p_{1})\}\text{.}
\label{ESScondsRSP}%
\end{align}
Suppose $(p^{\star},p_{1}^{\star})$ is a mixed NE then%

\begin{equation}
\left\{  \frac{\partial P}{\partial p}\mid_{\substack{p=q=p^{\star}%
\\p_{1}=q_{1}=p_{1}^{\star}}}(p^{\star}-p)+\frac{\partial P}{\partial p_{1}%
}\mid_{\substack{p=q=p^{\star}\\p_{1}=q_{1}=p_{1}^{\star}}}(p_{1}^{\star
}-p_{1})\right\}  \geq0\text{.}%
\end{equation}
Using substitutions%

\begin{equation}%
\begin{array}
[c]{cc}%
\left|  c_{11}\right|  ^{2}-\left|  c_{31}\right|  ^{2}=\bigtriangleup_{1}, &
\left|  c_{21}\right|  ^{2}-\left|  c_{11}\right|  ^{2}=\bigtriangleup_{1}^{%
\acute{}%
}\\
\left|  c_{13}\right|  ^{2}-\left|  c_{33}\right|  ^{2}=\bigtriangleup_{2}, &
\left|  c_{22}\right|  ^{2}-\left|  c_{12}\right|  ^{2}=\bigtriangleup_{2}^{%
\acute{}%
}\\
\left|  c_{12}\right|  ^{2}-\left|  c_{32}\right|  ^{2}=\bigtriangleup_{3}, &
\left|  c_{23}\right|  ^{2}-\left|  c_{13}\right|  ^{2}=\bigtriangleup_{3}^{%
\acute{}%
}%
\end{array}
\end{equation}
we get%

\begin{align}
\frac{\partial P}{\partial p}  &  \mid_{\substack{p=q=p^{\star}\\p_{1}%
=q_{1}=p_{1}^{\star}}}=p^{\star}(\bigtriangleup_{1}-\bigtriangleup
_{2})\left\{  (\alpha_{11}+\alpha_{33})-(\alpha_{13}+\alpha_{31})\right\}
+\nonumber\\
&  p_{1}^{\star}(\bigtriangleup_{1}-\bigtriangleup_{3})\left\{  (\alpha
_{11}+\alpha_{32})-(\alpha_{12}+\alpha_{31})\right\}  -\nonumber\\
&  \bigtriangleup_{1}(\alpha_{11}-\alpha_{31})-\bigtriangleup_{2}(\alpha
_{13}-\alpha_{33})-\bigtriangleup_{3}(\alpha_{12}-\alpha_{32})\text{,}
\label{RSPRestPoint1}%
\end{align}

\begin{align}
\frac{\partial P}{\partial p_{1}}  &  \mid_{\substack{p=q=p^{\star}%
\\p_{1}=q_{1}=p_{1}^{\star}}}=p^{\star}(\bigtriangleup_{3}^{%
\acute{}%
}-\bigtriangleup_{1}^{%
\acute{}%
})\left\{  (\alpha_{11}+\alpha_{23})-(\alpha_{13}+\alpha_{21})\right\}
+\nonumber\\
&  p_{1}^{\star}(\bigtriangleup_{2}^{%
\acute{}%
}-\bigtriangleup_{1}^{%
\acute{}%
})\left\{  (\alpha_{11}+\alpha_{22})-(\alpha_{12}+\alpha_{21})\right\}
+\nonumber\\
&  \bigtriangleup_{1}^{%
\acute{}%
}(\alpha_{11}-\alpha_{21})+\bigtriangleup_{2}^{%
\acute{}%
}(\alpha_{12}-\alpha_{22})+\bigtriangleup_{3}^{%
\acute{}%
}(\alpha_{13}-\alpha_{23})\text{.} \label{RSPRestPoint2}%
\end{align}
For the matrix (\ref{RSPMatrix}) the equations (\ref{RSPRestPoint1},
\ref{RSPRestPoint2}) can be written as%

\begin{align}
\frac{\partial P}{\partial p}  &  \mid_{\substack{p=q=p^{\star}\\p_{1}%
=q_{1}=p_{1}^{\star}}}=\bigtriangleup_{1}\left\{  -2\epsilon p^{\star
}-(3+\epsilon)p_{1}^{\star}+(1+\epsilon)\right\}  +\nonumber\\
&  \bigtriangleup_{2}\left\{  2\epsilon p^{\star}+(1-\epsilon)\right\}
+\bigtriangleup_{3}\left\{  (3+\epsilon)p_{1}^{\star}-2\right\} \\
\frac{\partial P}{\partial p_{1}}  &  \mid_{\substack{p=q=p^{\star}%
\\p_{1}=q_{1}=p_{1}^{\star}}}=\bigtriangleup_{1}^{%
\acute{}%
}\left\{  -p^{\star}(3-\epsilon)+2\epsilon p_{1}^{\star}+(1-\epsilon)\right\}
-\nonumber\\
&  \bigtriangleup_{2}^{%
\acute{}%
}\left\{  2\epsilon p_{1}^{\star}-(1+\epsilon)\right\}  +\bigtriangleup_{3}^{%
\acute{}%
}\left\{  (3-\epsilon)p^{\star}-2\right\}  \text{.}%
\end{align}
The payoff difference in the second condition of an ESS given in the Eq.
(\ref{ESScondsRSP}) reduces to%

\begin{align}
&  P\{(p^{\star},p_{1}^{\star}),(p,p_{1})\}-P\{(p,p_{1}),(p,p_{1}%
)\}\nonumber\\
&  =(p^{\star}-p)[-\bigtriangleup_{1}\{2\epsilon p+(3+\epsilon)p_{1}%
-(1+\epsilon)\}+\nonumber\\
&  \bigtriangleup_{2}\{2\epsilon p+(1-\epsilon)\}+\bigtriangleup
_{3}\{(3+\epsilon)p_{1}-2\}]+\nonumber\\
&  (p_{1}^{\star}-p_{1})[-\bigtriangleup_{1}^{%
\acute{}%
}\{(3-\epsilon)p-2\epsilon p_{1}-(1-\epsilon)\}-\nonumber\\
&  \bigtriangleup_{2}^{%
\acute{}%
}\{2\epsilon p_{1}-(1+\epsilon)\}+\bigtriangleup_{3}^{%
\acute{}%
}\{(3-\epsilon)p-2\}]\text{.}%
\end{align}
With the substitutions $(p^{\star}-p)=x$ and $(p_{1}^{\star}-p_{1})=y$ the
above payoff difference is%

\begin{align}
&  P\{(p^{\star},p_{1}^{\star}),(p,p_{1})\}-P\{(p,p_{1}),(p,p_{1}%
)\}\nonumber\\
&  =\bigtriangleup_{1}x\left\{  2\epsilon x+(3+\epsilon)y\right\}
-\bigtriangleup_{2}(2\epsilon x^{2})-\bigtriangleup_{3}xy(3+\epsilon
)-\nonumber\\
&  \bigtriangleup_{1}^{%
\acute{}%
}y\left\{  2\epsilon y-(3-\epsilon)x\right\}  +\bigtriangleup_{2}^{%
\acute{}%
}(2\epsilon y^{2})-\bigtriangleup_{3}^{%
\acute{}%
}xy(3-\epsilon) \label{2ndESSRSP}%
\end{align}
provided that%

\begin{equation}
\frac{\partial P}{\partial p}\mid_{\substack{p=q=p^{\star}\\p_{1}=q_{1}%
=p_{1}^{\star}}}=0\text{ \ \ \ \ \ \ \ \ \ \ \ \ \ }\frac{\partial P}{\partial
p_{1}}\mid_{\substack{p=q=p^{\star}\\p_{1}=q_{1}=p_{1}^{\star}}}=0\text{.}
\label{CondsRSP}%
\end{equation}
The conditions in Eq. (\ref{CondsRSP}) together define the mixed NE
$(p^{\star},p_{1}^{\star})$. Consider now the modified RSP game in classical
form obtained by setting $\left|  c_{11}\right|  ^{2}=1$. The Eqs.
(\ref{CondsRSP}) become%

\begin{align}
-2\epsilon p^{\star}-(\epsilon+3)p_{1}^{\star}+(\epsilon+1)  &  =0\nonumber\\
(-\epsilon+3)p^{\star}-2\epsilon p_{1}^{\star}+(\epsilon-1)  &  =0
\end{align}
and $p^{\star}=p_{1}^{\star}=\frac{1}{3}$ is obtained as a mixed NE for all
the range $-1<\epsilon<0$. From Eq. (\ref{2ndESSRSP}) we get%

\begin{align}
&  P\{(p^{\star},p_{1}^{\star}),(p,p_{1})\}-P\{(p,p_{1}),(p,p_{1}%
)\}\nonumber\\
&  =2\epsilon(x^{2}+y^{2}+xy)=\epsilon\left\{  (x+y)^{2}+(x^{2}+y^{2}%
)\right\}  \leq0\text{.} \label{differ}%
\end{align}
In the classical RSP game, therefore, the mixed NE $p^{\star}=p_{1}^{\star
}=\frac{1}{3}$ is a NE but not an ESS, because the second condition of an ESS
given in the Eq. (\ref{ESScondsRSP}) does not hold.

Now define a new initial state as%

\begin{equation}
\left|  \psi_{in}\right\rangle =\frac{1}{2}\left\{  \left|  12\right\rangle
+\left|  21\right\rangle +\left|  13\right\rangle +\left|  31\right\rangle
\right\}  \label{InistatRSP}%
\end{equation}
and use it to play the game, instead of the classical game obtained from
$\left|  \psi_{in}\right\rangle =\left|  11\right\rangle $. The strategy
$p^{\star}=p_{1}^{\star}=\frac{1}{3}$ still forms a mixed NE because the
conditions (\ref{CondsRSP}) hold true for it. However the payoff difference of
Eq. (\ref{2ndESSRSP}) is now given below, when $-1<\epsilon<0$ and $x,y\neq0$:%

\begin{align}
&  P\{(p^{\star},p_{1}^{\star}),(p,p_{1})\}-P\{(p,p_{1}),(p,p_{1}%
)\}\nonumber\\
&  =-\epsilon\left\{  (x+y)^{2}+(x^{2}+y^{2})\right\}  >0\text{.}%
\end{align}
Therefore, the mixed NE $p^{\star}=p_{1}^{\star}=\frac{1}{3}$, not existing as
an ESS in the classical form of the RSP game, becomes an ESS when the game is
quantized and played using an initial (entangled) quantum state given by the
Eq. (\ref{InistatRSP}).

Note that from Eq. (\ref{3StrategyPayoff}) the sum of the payoffs to Alice and
Bob $(P_{A}+P_{B})$ can be obtained for both the classical mixed strategy game
(i.e. when $\left|  \psi_{in}\right\rangle =\left|  11\right\rangle $) and the
quantum game played using the quantum state of Eq. (\ref{InistatRSP}). For the
matrix (\ref{RSPMatrix}) we write these sums as $(P_{A}+P_{B})_{cl}$ and
$(P_{A}+P_{B})_{qu}$ for classical mixed strategy and quantum games,
respectively. We obtain%

\begin{equation}
(P_{A}+P_{B})_{cl}=-2\epsilon\left\{  (1-p-p_{1})(1-q-q_{1})+p_{1}%
q_{1}+pq\right\}
\end{equation}
and%

\begin{equation}
(P_{A}+P_{B})_{qu}=-\left\{  \frac{1}{2}(P_{A}+P_{B})_{cl}+\epsilon\right\}
\text{.}%
\end{equation}
In case $\epsilon=0$ both the classical and quantum games are clearly zero
sum. For the slightly modified version of the RSP game we have $-1<\epsilon<0$
and both versions of the game become non zero-sum.

\section{Concluding Remarks}

Evolutionary stability is a game-theoretic solution concept that tells which
strategies are going to establish themselves in a population of players
engaged in symmetric contests. By establishing itself it means that the
strategy becomes resistent to invasion by mutant strategies when played by a
small number of players. Analysis of evolutionary stability in quantum games
shows that quantization of games, played by a population of players, can lead
to new stable states of the population in which, for example, a quantum
strategy establishes itself. Our results show that quantum strategies
\index{Quantum strategy}
can indeed change the dynamics of evolution as described by the concept of
evolutionary stability. Quantum strategies
\index{Quantum strategy}
being able to decide evolutionary outcomes clearly gives a new role to quantum
mechanics
\index{Quantum mechanics}
which is higher than just keeping the atoms together. Secondly, evolutionary
stability in quantum games provides a mathematically tractable method of
analysis for studying multi-player quantum games
\index{Multi-player quantum games}
\cite{Benjamin Multiplayer} played in evolutionary arrangements.

Using EWL and MW quantization schemes
\index{EWL quantization scheme}
\index{MW quantization scheme}%
, we explored how quantization can change evolutionary stability of Nash
equilibria in certain asymmetric bi-matrix games. The ESS concept was
originally defined for symmetric bi-matrix contests. We showed that
quantization can change evolutionary stability of a NE also in certain types
of symmetric bi-matrix games. We identified the classes of games, both
symmetric and asymmetric, for which within the EWL
\index{EWL quantization scheme}
and MW schemes
\index{MW quantization scheme}
the quantization of games becomes related to evolutionary stability of NE. For
example, in the case of Prisoners' Dilemma we found that when a population is
engaged in playing this symmetric bi-matrix game, a small number of mutant
players can invade the classical ESS\footnote{consisting of
Defection-Defection strategy pair} when they exploit Eisert and Wilken's
two-parameter set of quantum strategies. As another example we studied the
well-known childrens' two-player three-strategy game of Rock-Scissors-Paper.
In its classical form a mixed NE exists that is not evolutionarily stable. We
found that in a quantum form of this game, played using MW quantization
scheme
\index{EWL quantization scheme}%
, the classical NE becomes evolutionarily stable when the players share an
entangled state.

We speculate that evolutionary stability in quantum games can potentially
provide a new approach towards the understanding of rise of complexity and
self-organization
\index{Complexity and self-organization}
in groups of quantum-interacting entities, although this opinion, at the
present stage of development in evolutionary quantum game theory
\index{Evolutionary quantum game theory}%
, remains without any supportive evidence, either empirical or experimental.
However, it seems that the work presented in this chapter provides a
theoretical support in favor of this opinion. Secondly, evolutionary quantum
game theory benefits from the methods and concepts of quantum mechanics
\index{Quantum mechanics}
and evolutionary game theory, the second of which is well known to facilitate
better understanding of complex interactions taking place in communities of
animals as well as that of the bacteria and viruses. Combining together the
techniques and approaches of these two, seemingly separate, disciplines
appears to provide an ideal arrangement to understand the rise of complexity
and self-organization
\index{Complexity and self-organization}
at molecular level.

Although it is true that evolutionary stability and evolutionary computation
provide two different perspectives on the dynamics of evolution, it appears to
us that an evolutionary quantum game-theoretic approach can potentially
provide an alternative viewpoint in finding evolutionary quantum search
algorithms that may combine the advantages of quantum and evolutionary
computing
\index{Evolutionary computing}
\cite{Greenwood}. This will then also provide the opportunity to combine the
two different philosophies representing these approaches towards computing:
evolutionary search
\index{Evolutionary search}
and quantum computing.

%

\printindex

Biographies:

\medskip

\emph{Azhar Iqbal}\textit{ graduated in Physics in 1995 from the University of
Sheffield, UK. From 1995 to 2002 he was associated with the Pakistan Institute
of Lasers \& Optics. He earned his PhD from the University of Hull, UK, in
2006 in the area of quantum games. He is Assistant Professor (on leave) at the
National University of Sciences and Technology, Pakistan and Visiting
Associate Professor at the Kochi University of Technology, Japan.}

\medskip

\emph{Taksu Cheon}\textit{ graduated in Physics in 1980 from the University of
Tokyo, Japan. He earned his PhD from the University of Tokyo in 1985 in the
area of theoretical nuclear physics. He is Professor of Theoretical Physics at
the Kochi University of Technology, Japan.}

\end{document}